\documentclass[12pt,preprint]{aastex} 
\usepackage{amssymb, amsmath,fullpage,bm} 
\usepackage{lscape}
\usepackage[dvips]{color}
\usepackage[usenames,dvipsnames]{xcolor}
\usepackage{color}
\shorttitle{Reproducing the abundances in RCB stars with post-DD models}
\shortauthors{Menon et al.}

\usepackage{ifpdf}

\ifpdf

\usepackage[pdftex]{epsfig}

\usepackage[pdftex]{hyperref}

\else

 \usepackage[dvips]{epsfig}

\newcommand{\href}[2]{#2}

\fi

\begin{document}
\title{Reproducing the observed abundances in RCB and HdC stars with
  post-double degenerate merger models - constraints on merger and
  post-merger simulations and physics processes}

\author{Athira Menon\altaffilmark{1}, Falk Herwig\altaffilmark{1,2},
  Pavel A. Denissenkov\altaffilmark{1,2}}
\affil{Department of Physics \& Astronomy, University of Victoria, Victoria,
BC V8P5C2, Canada}
\author{Geoffrey C. Clayton\altaffilmark{1},  Jan Staff}
\affil{Department of Physics and Astronomy, Louisiana State University,
202 Nicholson Hall, Tower Dr., Baton Rouge, LA 70803-4001,
USA}
\author{Marco Pignatari\altaffilmark{1}}
\affil{Department of Physics, University of Basel,
Klingelbergstrasse 82, CH-4056 Basel, Switzerland}
\author{Bill Paxton}
\affil{Kavli Institute for Theoretical Physics and Department of
  Physics, Kohn Hall, University of California, Santa Barbara, CA
  93106, USA}

\altaffiltext{1}{NuGrid collaboration, \url{http://www.nugridstars.org}}
\altaffiltext{2}{The Joint Institute for Nuclear Astrophysics}


%
%

\begin{abstract} 
  The R Coronae Borealis (RCB) stars are hydrogen-deficient, variable
  stars that are most likely the result of He-CO WD mergers. They
  display extremely low oxygen isotopic ratios, $^{16}$O/$^{18}$O
  $\simeq 1 - 10 $, $^{12}\mathrm{C}/^{13}\mathrm{C}\geq 100$, and
  enhancements up to $2.6 \mathrm{dex}$ in F and in s-process elements
  from Zn to La, compared to solar.  These abundances provide
  stringent constraints on the physical processes during and after the
  double-degenerate merger.  As shown before O-isotopic ratios
  observed in RCB stars cannot result from the dynamic
  double-degenerate merger phase, and we investigate now the role of
  the long-term 1D spherical post-merger evolution and nucleosynthesis
  based on realistic hydrodynamic merger progenitor models.  We adopt
  a model for extra envelope mixing to represent processes driven by
  rotation originating in the dynamical merger.  Comprehensive
  nucleosynthesis post-processing simulations for these stellar
  evolution models reproduce, for the first time, the full range of
  the observed abundances for almost all the elements measured in RCB
  stars: $^{16}\mathrm{O}$/$^{18}\mathrm{O}$ ratios between 9 and 15,
  C-isotopic ratios above 100, and
  $\sim$\,1.4\,--\,2.35\,$\mathrm{dex}$ F enhancements, along with
  enrichments in s-process elements. The nucleosynthesis processes in
  our models constrain the length and temperature in the dynamic
  merger shell-of-fire feature as well as the envelope mixing in the
  post-merger phase. s-process elements originate either in the
  shell-of-fire merger feature or during the post-merger evolution,
  but the contribution from the AGB progenitors is negligible. The
  post-merger envelope mixing must eventually cease $\sim
  10^{6}\mathrm{yr}$ after the dynamic merger phase, before the star
  enters the RCB phase.
\end{abstract}


%
%
\section{Introduction}
\label{sec:intro}
RCB stars are rare F-G type supergiants belonging to the old bulge
population in the Galaxy \citep{cottrell98,tisserand08}. Their
atmospheres are mainly composed of He, almost entirely depleted
in H yet rich in C. They are near-solar mass single stars \citep[except for
DY Cen which is in a binary system,][]{rao12} that have single or
multimode pulsation cycles. The pulsations are considered to play a
role in the deep, irregular brightness declines of up to 8 magnitudes
\citep{clayton96}. These declines are attributed to clouds of dust
ejected from their atmospheres, the evidence for which is an
omnipresent IR signature around the star \citep{clayton96}. Such deep
declines and IR excesses have not been observed in the otherwise
chemically similar Hydrogen-deficient Carbon (HdC) stars or the hotter
Extreme Helium (EHe) stars.

Other than the declines for which they were first noticed, RCB stars
have been an enigma because of their anomalous chemical signatures and
the possible evolutionary path (or paths) that may have led to their
current state. The surface layers of RCBs are predominantly made of helium
(mass fraction $X_\mathrm{He}$=0.98) and are carbon rich with
C$>$N$>$O. They also have $^{12}$C/$^{13}$C ratios larger than 40\,--\,100
\citep{hema12} (compare with the CN-cycle equilibrium value $\approx$\,3.5 
that is expected for H-burning ashes), F enhancements
in the range of 1.0\,--\,2.7\,$\mathrm{dex}$ \citep{pandey08} and
considerable overabundances of s-process elements compared to solar
values \citep{asplund00}. This peculiar surface composition is indicative of
material processed by H-burning and (partial) He-burning. The
$^{16}\mathrm{O} / ^{18}\mathrm{O}$ ratios in HdC and RCB stars have been a
focus of many recent studies, starting with the discovery of an
extremely low value of 0.83$\pm$0.02 in the HdC star HD 137613
\citep{clayton05}, as compared to the solar value of 500. RCB stars, as a
class, have $^{16}\mathrm{O}$/$^{18}\mathrm{O}$ ratios between 1 and 20
\citep{clayton07,garcia10}, probably the lowest ones measured anywhere in
the Universe. 
The goal of this work is to construct post-merger models that can
reproduce the unique chemical signatures of RCB stars,
particularly their low $^{16}\mathrm{O}$/$^{18}\mathrm{O}$ ratios and F enhancements.

The EHe stars have similar compositions (barring a few exceptions) as
RCB stars, suggesting a common evolutionary origin and link. The surfaces
of EHe stars are too hot to show CO bands, therefore their
$^{16}\mathrm{O}$/$^{18}\mathrm{O}$ ratios cannot be measured. As one
of our primary goals in this paper is to model the low $^{16}\mathrm{O}$/$^{18}\mathrm{O}$
ratios seen in RCB stars, we exclude EHe stars from our discussion. It
should be noted however that EHes do have high F enhancements
reported, like those in RCB stars \citep{pandey08}.


65 RCB stars have been found in the Milky Way and 25 in the Magellanic
Clouds \citep{clayton12}. Two competing scenarios
have been put forth to explain how RCB stars may have formed - the
Final He-Flash (FF) model \citep{renzini90} and the Double Degenerate
(DD) white dwarf merger model \citep{webbink84,iben96}. In the FF model, a
post-AGB star experiences a He-shell flash during the pre-white dwarf
phase and evolves back into a giant configuration. In the DD merger
scenario, a CO white dwarf and a less massive He white dwarf (WD) in a
close binary system, formed as the result of one or more common envelope
episodes, finally merge over a dynamic period of $\approx$\,100\,--\,1000\,$\mathrm{s}$. 
In terms of number statistics, the expected number of RCB
stars resulting from the merger of CO and He WDs is very close to the
estimated $5700$ RCB stars obtained by extrapolating the LMC RCB
population \citep{clayton12}, which implies that only a small fraction
has been found so far. Computational models and observational evidence
also lean toward the DD model being the more plausible path out of the
two scenarios to form RCB stars
\citep{clayton07,loren-aguilar09,garcia09}. The main nucleosynthesis
evidence comes from the $^{12}$C/$^{13}$C and
$^{16}\mathrm{O}$/$^{18}\mathrm{O}$ ratios in RCB stars.

According to a single-zone analysis, rapid mass-transfer events in
He+CO WD binaries could lead to thermodynamic conditions in which such
ratios would originate \citep{clayton07}.  Specifically,
$^{18}\mathrm{O}$ can be produced by partial He-burning via
$^{14}\mathrm{N(\alpha,\gamma)}^{18}\mathrm{F(\beta^+)}^{18}\mathrm{O}$,
without being destroyed by a further $\alpha$-capture to
$^{22}\mathrm{Ne}$ for a temperature in a range of 1.2\,--\,1.9$\times
10^{8}\,\mathrm{K}$ at a density of $10^{3} \mathrm{g/cm^{3}}$ and on time
scales of $\approx$\,250\,--\,320\,$\mathrm{yr}$. At these temperatures,
$^{13}\mathrm{C}$ is consumed by He-burning to form $^{16}\mathrm{O}$
via $^{13}\mathrm{C(\alpha,n)}^{16}\mathrm{O}$, which leads to the high
C-isotopic ratios observed in RCB stars, while the neutrons released
from this reaction could produce s-process elements.


The thermal properties of a final-flash event are equivalent to those of an AGB
thermal pulse, with temperatures in the convective He-burning shell as
high as $3 \times 10^{8} \,\mathrm{K}$ \citep[][and ref.\
therein]{werner:06,herwig11} maintained for a long enough time to process
$^{18}\mathrm{O}$ and $^{19}\mathrm{F}$ to $^{22}\mathrm{Ne}$ (the
latter via the reaction
$^{19}\mathrm{F(\alpha,p)}^{22}\mathrm{Ne}$). This may explain the
lack of $^{18}\mathrm{O}$ \citep[from CO bands,][]{eyres98,geballe02})
and F\,I bands \citep{pandey08} in the spectrum of Sakurai's object
(V4334 Sgr) which is believed to be a final-flash star of the very
late thermal pulse (VLTP) variant \citep[][and references therein]{herwig:01ApJ}. 
In the VLTP event, $^{13}$C is produced by
proton ingestion into the convection zone and their capture by $^{12}$C
resulting in low $^{12}$C/$^{13}$C ratios of $\sim$\,5
\citep{herwig01,herwig11}, which compares well with the observed value
of $1.5-5$ in Sakurai's object \citep{asplund97,asplund99}. VLTP star
models also predict Li enrichments \citep{herwig01}, as observed in
Sakurai's object, but there are only four Li-rich RCB stars.

A compact system of two WDs evolves from a pair of main
sequence stars that are close enough to interact with each other
through the exchange of gravitational wave radiation and magnetic
stellar winds, causing a loss of orbital angular momentum from the
system and thus reducing the orbital separation between the
stars. Such close double-degenerate WD systems form over a period of
$\sim$\,$10^{9}$\,years \citep{iben84,iben87} through one or more
episodes of mass-transfer (including at least one common-envelope
interaction). The final merging of the two WDs occurs on a dynamic
timescale (100\,--\,1000 seconds), with the less massive WD (the He WD)
being tidally disrupted by the more massive one (the CO WD). Smooth
Particle Hydrodynamic (SPH) simulations that follow this merging phase
\citep{loren-aguilar09,raskin12,staff12} show that the He WD material rapidly
infalling on the CO WD causes the surface of the latter to
be shock-heated, resulting in a hot corona feature, while the rest of
the He-WD mass settles in a disk-like structure that extends to
large radii.

One-dimensional He-rich homogeneous stellar evolution models with C, O
and trace amounts of H showed promising evolutionary tracks into the
RCB region in the HR diagram \citep{weiss87}.  In an alternative
approach, \citet{saio02} represented the merger phase through accreting
He-WD material (of a pure helium composition with a small H-rich
envelope mass) onto a CO WD at the rate of $10^{-5}
M_\mathrm{{\odot}}/$year. Although the location of the final model in
the HR diagram was consistent with that of EHes and RCB stars, the
accretion rate was much smaller than the expected rapid mass-transfer
rate of $\sim$\,$10^{3} M_\mathrm{{\odot}}/$year
\citep{staff12,loren-aguilar09} during the final merging phase of
WDs.

The SPH simulations of the dynamic merger phase of He and CO WDs,
which were performed over a range of mass combinations by
\citet{loren-aguilar09}, included a 14-species nuclear network.  The
abundance yields in the hot corona of the final models of masses $\geq$\,$1.2\,M_\mathrm{\odot}$ 
showed enrichments of Ca, Mg, S, Si, and
Fe. \citet{longland11} carried out a one-zone
post-processing nucleosynthesis analysis using the temperature and density conditions
from the corona of the final model. The AGB model used to build their
initial nuclear composition contained 7 nuclear species from
$^{1}\mathrm{H}$ to $^{22}\mathrm{Ne}$. They reported that, depending on
the depth of mixing within the corona, the
$^{16}\mathrm{O}$/$^{18}\mathrm{O}$ ratio had decreased to $\approx$\,19\,--\,370. However, their
total C and O abundances exceeded the upper limit for RCB stars and
their F predictions were much lower than the observed values, while predictions for
s-process element production were not available.

\citet{jeffery11} have investigated RCB surface abundances by using a
cold mixing recipe, i.e.\ without nucleosynthesis during either the
hot merger phase or the post-merger evolution. A 1-D `cold merger'
model involved mixing different proportions of masses of He
and CO WDs, and a `hot merger' model was built by incorporating the
results obtained in the SPH simulations of He-CO WD merger systems by
\citet{loren-aguilar09}. In both cases, 
the elemental abundances obtained from their mixing recipe,
barring those of C, N and O (and S in the hot merger model), 
fell short of the observed ones in RCB stars,
particularly the s-process elements, which were assumed to originate
exclusively from the progenitor AGB star. They  concluded that even
additional nucleosynthesis that may occur during the actual merger of
He and CO WDs cannot help to reproduce the observed abundances of
RCB stars.
 
Continuing the work by \citet{clayton05} and \citet{clayton07},
\citet{staff12} (paper I) presented grid-based hydrodynamic merger
simulations for a range of mass ratios ($q$) of CO and He WDs (of total
mass $0.9\,M_\mathrm{\odot}$). A Shell of Fire (SOF)
\citep[similar to the hot corona of][]{loren-aguilar09} was identified
in the merged object, and that feature was analyzed for the potential
formation of low $^{16}\mathrm{O}$/$^{18}\mathrm{O}$ ratio
material. Single-zone nucleosynthesis calculations using the $T$,
$\rho$ conditions of the SOF of three cases were made for a
range of initial abundance distributions derived from a variety of
appropriate AGB progenitor models. Those were taken from the
NuGrid\footnote{\url{http://www.nugridstars.org}} library of detailed
post-processed, fully evolved models of solar metallicity that ranged
from the early AGB to the advanced thermal pulse stage (Set 1.2,
Pignatari et al., in prep). The composition of the He WD was taken to
be uniformly mixed and was extracted from a low mass RGB star. In
contrast to the results of \citet{longland11}, we found that within
the dynamic phase, the $^{16}\mathrm{O}$/$^{18}\mathrm{O}$ ratio in
the SOF was of the order of 1000. The difference in the
$^{16}\mathrm{O}$/$^{18}\mathrm{O}$ ratios between the two simulations may
arise from the difference in the total mass considered. In any
case, according to our results, the low observed
$^{16}\mathrm{O}$/$^{18}\mathrm{O}$ ratios of RCB stars cannot 
originate from the dynamic merger phase, but if the temperature and
density conditions were maintained for a longer period of time,
then $^{16}\mathrm{O}$/$^{18}\mathrm{O}$ ratios between 4 ($q=0.7$,
$10^{9}\,\mathrm{s}$) to 30 ($q=0.5$, $10^{6}\,\mathrm{s}$) could be
obtained.

Following the results of paper I, we now explore the the long term
post-merger evolution with the goal of identifying conditions under which
this evolutionary phase could reconcile the present stark discrepancy
between models and observed surface abundance of RCB stars.  We are
motivated by the fact that merging WDs are considered to be paths to the
creation of many other interesting stellar objects,  such as Type Ia
supernovae and neutron stars \citep{han98,nel01a,nap01}, sdB stars
\citep{saio00,han02}, AM CVns \citep{solheim10} and even planetary
nebulae \citep{demarco09}. The rich and detailed observational picture
available for RCB and HdC stars gives an opportunity to constrain
DD merger and post-merger physics in general, thereby providing a
quantitative understanding of the objects mentioned above.

The long term evolution of such merged systems may be characterized by
the formation of an accretion disk from the secondary WD
\citep{yoon07,van10}.  Alternatively, the post-merger object may
undergo a more viscous, star-like evolution with the Keplerian rotation in the
envelope eventually dying out \citep{shen12,schwab12}. We follow the
latter scenario for the post-merger evolution, and we assume the
merged object to evolve as a hybrid C/O-He star undergoing
rotation-driven  mixing in its envelope. We construct
spherically symmetric one-dimensional models of merged WDs, with a
range of initial conditions and follow their evolution into the domain
of the HR diagram where RCB stars are observed to lie. Along with
convective mixing, these models also include an induced, continuous
mixing profile assumed to result from the rotation of the
merger-remnant. We then perform a multi-zone post-processing
nucleosynthesis analysis of these models and compare their surface
abundances with those of RCB stars.

This paper is organized as follows. Section~2 describes our simulation methods,
where the progenitor evolution of the system, the construction of the
initial models, a brief description of the codes we used, the formulation
of the mixing profile, and the four post-merger evolution cases that we studied.
This is followed by Section~3, where the stellar and
nucleosynthesis processes of the model during its evolution are
described and compared with each other and with the RCB star
observations. We close with the discussion and conclusions in Section~4.

%
%
\section{Methods}

The post-merger simulations are building on our previous dynamic
merger simulations (paper I). They involve the construction of
appropriate initial models for the post-merger evolution, starting
with the abundance profiles that take into account information for all
observed species from realistic progenitor models as well as our best
estimates of nucleosynthesis in the dynamic merger phase (i.e.\ in the
SOF). The initial abundance profiles are homologously imposed on a
homogeneous zero-age main-sequence model  of the same mass, so that
within a thermal time scale the stellar structure is determined by and
consistent with the post-merger abundance profile. The stellar
evolution of these objects is followed into the RCB region of the HRD,
and then post-processed with the same complete nuclear network (using
the NuGrid PPN codes) that has been used to generate the progenitor
models. In this way our models take into account nuclear production
during the progenitor evolution, the hot merger phase (paper I) as
well as any production during the post-merger phase, which - as we
show here - is essential to reproduce the observed abundances of RCB
stars.

\subsection{Progenitor evolution and outcome of dynamic merger phase}

The progenitor evolution of a double-degenerate merger evolves from a
close binary of low- and intermediate-mass main sequence stars, and involves a
common envelope (CE) event when the primary becomes a giant star
\citep[e.g.][]{iben96}. CE events are expected to have occurred in
close binary systems with at least one WD component
\citep{demarco11,passy12}. The outcome of the second CE event is a
close binary with CO and He WDs. This progenitor evolution
matters here only in as much as it determines which RGB and AGB models
should be considered when constructing the initial abundance
distribution out of their combined He and CO cores, as described in
\citet{staff12}.

The final merging of the two WDs takes place within a dynamic
timescale of $10^{2}$\,--\,$10^{3}\,\,\mathrm{s}$ and results in a structure
consisting of a CO core surrounded by a shock-heated layer and an
envelope in Keplerian rotation
\citep{loren-aguilar09,dan12,raskin12}. In our hydrodynamic
simulations reported in paper I, this hot and very dense SOF lasted
until the end of the simulations. The SOF is the
region where nucleosynthesis can occur within the dynamic time scale.
The final nuclear production of the SOF will depend on how
long it lasts beyond the end of simulations. 

The post-merger evolution is determined by the interplay of thermal
dissipation, angular momentum transport induced by rotational
instabilities and magnetic fields, and the nuclear time scale of H-
and He-burning shells. How these ingredients interact is still
uncertain, and with regard to the influence that nuclear burning has
on this phase differences may exist depending on the composition of the
merging WDs. Considering the merger of two CO WDs the merger has been
described as an accretion of the post-merger disk remnant onto the
core over a timescale of either $10^{5}$\,years \citep{yoon07} or few
hours \citep{van10}. Alternatively, the post-merger object couples the
rigid primary WD and the tidally disrupted secondary WD undergoing
Keplerian rotation through viscous mechanisms (from
magnetohydrodynamic instabilities) that stimulate transport of angular
momentum over a period between $10^{4}$ and $10^{8}$ seconds
\citep{shen12}. These shearing forces may also increase the peak
temperature of the SOF/corona of the remnant by a factor of
$\approx$\,2 and work on expanding the material to larger
radii. \citet{schwab12} considered a range of WD combinations
including He+CO WD mergers. They show that, towards the end of such a
viscous phase, the remnant evolves towards a spherically symmetric,
almost shear-free steady state with a thermally-supported envelope in
solid body rotation (although the case with the smallest CO to He WD
mass ratio shows the largest asymmetries of all cases). These authors
expect the subsequent evolution to be star-like, growing into a giant
star, with the luminosity from its internal nuclear-burning driving
convection in its cool extended envelope. Such evolution represents a
path from the merger of white dwarfs to the formation of RCB giant
stars.

Following the latter scenario, we construct the initial composition
profile for our long-term post-merger simulations to be comprised of a
CO core, an SOF, and an envelope with rotationally induced mixing,
predominantly made of the former He WD. 

\subsubsection{The post-merger composition profile}
\label{sec:initial_comp}
The first step in building the internal composition of the post-merged
object is to choose the progenitor AGB and RGB stars of the CO and He
WDs respectively. These progenitor models determine when
the CE interactions will take place and therefore cannot be uniquely
determined {\em a posteriori}. We consider two out of the three CO-WD progenitor
models introduced in paper I. Accordingly, the CO WD is the progeny of
either an early-AGB (E-AGB) or advanced-thermal pulse AGB (A-TP AGB)
star. The difference between these progenitors is in the amount of
s-process elements in the outer layers of the CO WD\footnote{However
  we find that this s process material in the progenitor AGB is not
  the origin of the observed s-process overabundances in RCB stars,
  cf.\ Section~\ref{sec:results}}, as well as the
thickness of the He-rich layer. In both cases, the CO WD has a H-rich
envelope of the order of $10^{-4}$$\,M_\mathrm{\odot}$. The RGB model
for the He WD has a more massive H-rich envelope of
$\sim$\,$10^{-3}$$\,M_\mathrm{\odot}$. These WD mass dependent
amounts of H have been chosen according to stellar evolution
simulation results \citep[see Fig.\,8, paper I]{staff12} and play - as we
will show - an important role in reproducing both the CNO elemental
abundances as well as providing for the formation of the neutron
source isotope $^{13}\mathrm{C}$.  The masses of the different layers of the
WDs are listed in Table~\ref{nuc_mod}.

\begin{deluxetable}{ccccc}
\tablecolumns{5} \tabletypesize{\footnotesize}
\tablecaption{The WD masses used for the initial models of the post-merger
  evolution. $M_\mathrm{WD-core}$ is the mass of the WD core (Fig.\,\ref{ini_profile}), while
  $M_\mathrm{env}$ is the mass of the H/He- or H-envelope for the CO
  and He WDs, respectively.}
\tablewidth{0pt}
\tablehead{
\colhead{WD} & \colhead{Progenitor ($M_{\rm init}/M_\mathrm{\odot}$)} &
\colhead{$M_\mathrm{WD}$/$\,M_\mathrm{\odot}$} &
\colhead{$M_\mathrm{WD-core}$/$\,M_\mathrm{\odot}$} & \colhead{$M_\mathrm{env} /10^{-4}$$\,M_\mathrm{\odot}$}}
\startdata
CO & E-AGB (3.00) & 0.58148 & 0.4876 (He-free) & 2.4 \\
CO & A-TP AGB (2.00) & 0.61243 & 0.5789 (He-free) & 1.7 \\
He & RGB (1.65) & 0.3024 & 0.296 (H-free) & 64.3 \\
\enddata
\label{nuc_mod}
\end{deluxetable}

\begin{figure}
\includegraphics[scale=0.80]{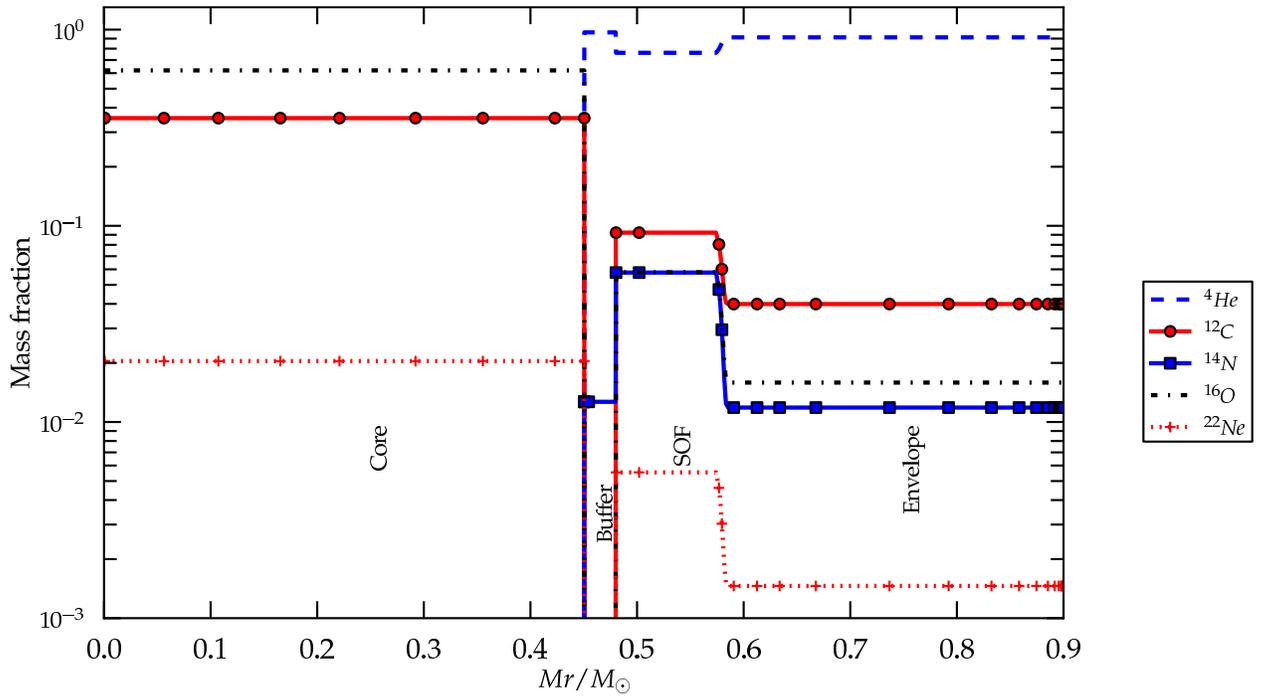}
\caption{Four-zone post-merger abundance profiles of the $q=$0.7, E-AGB CO WD with a dredge-up depth of 15$\%$ (case 1).}
\label{ini_profile}
\end{figure}

The various components of the pre-merger WD composition are combined
by taking into account the dynamic mixing and the SOF nuclear burning,
both derived from the merger simulations of paper I. This
leads to a four-zone model consisting of the
relatively cold, degenerate core of the CO WD, a thin buffer zone, the
SOF itself, and a relatively cold envelope
(Fig.\,\ref{ini_profile}). 

For each mass ratio, $q=M_\mathrm{ He}/M_\mathrm{CO}$, a certain
fraction of mass ($f_\mathrm{dre}$) is dredged up from the CO WD
($M_\mathrm{CO}$) and mixed with the He WD ($M_\mathrm{ He}$) material
in the envelope. The amount of dredge-up defines the outer boundary of
the core, $m_\mathrm{core}$ (Eqs.~\ref{eq:mass_eq}). As this
dredged-up mass from the He-rich layers of the CO WD mass ($m_\mathrm{
  dre}$) is pulled up to the envelope \citep[see Fig.\,11 of paper I ]{staff12},
an equal mass of He WD makes its way into the CO WD. Thus a
partially mixed zone is formed between the two WDs. A major portion of
the mass in this zone is occupied by the SOF ($m_\mathrm{SOF}$), while the
rest is in the buffer zone between the core and the SOF
($m_\mathrm{buffer}$). A small fraction of the mass $m_\mathrm{ dre}$
from the He WD (that penetrates the CO WD and forms the partial mixing
zone) resides in the SOF ($m_\mathrm{ he-sof}$), while a larger part
finds itself located in the buffer zone\footnote{The detailed
  properties of this buffer zone are difficult to extract from the 3D
  merger simulations for the purpose of mapping them into the 1D
  profile. However, it turns out that these details are not important
  for the final RCB abundance predictions.}.

The initial SOF composition is thus dominated by the mass dredged up
from the CO WD ($m_\mathrm{ co-sof}$). As in paper I, we assume that
the exchange of masses during the dynamic merger phase occurs before
the onset of burning in any given layer. Correspondingly, the SOF layer is
subjected to nuclear processing according to the conditions in the SOF
derived from the merger calculations (see Section~\ref{sec:cases} for
two cases considered for SOF processing). The result of this nuclear
processing is then entered into the post-merger abundance
profile. Finally, the envelope ($m_\mathrm{ env}$) contains the
fraction of $m_\mathrm{ dre}$ from the CO WD that is not present in
the SOF and the remaining He WD mass.
\begin{eqnarray}
\label{eq:mass_eq}
\nonumber
m_\mathrm{ dre} = f_\mathrm{ dre} \times  M_\mathrm{ CO}  \\
\nonumber
m_\mathrm{ core} = M_\mathrm{ CO} - m_\mathrm{ dre} \\
m_\mathrm{ buffer} =  m_\mathrm{ dre} - m_\mathrm{ he-sof} \\
\nonumber
m_\mathrm{ SOF} = m_\mathrm{ he-sof} + m_\mathrm{ co-sof} \\
\nonumber
m_\mathrm{ env} = M_\mathrm{ He} - m_\mathrm{ co-sof}\\
\nonumber
\end{eqnarray} 

Motivated by the merger simulations, we assume that the He-rich
envelope is completely mixed during the merger process. The most
important consequence of this assumption is the uniform distribution
in the post-merger envelope of H from the former H-rich He-WD
envelope. After extracting the abundances from each zone of the
progenitor models, the isotopic abundances of that zone are averaged
over its mass.
 
\subsubsection{The four post-merger cases} 
\label{sec:cases}
The key parameters of the post-merger models are the pre-merger WD
mass ratio, the amount of
dynamic dredge-up from the CO WD, the type of CO-WD progenitor model,
and the temperature and duration of the SOF nuclear burning in the
dynamic phase. We focus here on four representative combinations of
these parameters.

The dynamic merger simulations predict dredge-up depths ranging from
15{\%} to 18{\%} of mass into the CO WD for the three low-$q$
cases. In the outer layers of the He-free core of the CO WD
$^{16}\mathrm{O}$ is the most abundant isotope. If material is
dredged up from these layers the star becomes more enhanced in oxygen
than in carbon at its surface, which is not observed in RCB stars. The
other problem of dredging up high amounts of oxygen is that the star
would have much higher $^{16}\mathrm{O} / ^{18}\mathrm{O}$ ratios than
observed in RCB stars, even when post-merger nucleosynthesis effects are
taken into account (see below). Therefore, in order to avoid
an excessive surface enrichment in $^{16}\mathrm{O}$, the adopted mass of dredge-up
from the chosen CO WD models is lower than the values
obtained in the hydrodynamic simulations. We thus restrict the dredge-up
depth in our models to layers just above the He-free core of the CO WD
(ranging from $\sim$\,5.4\% in the A-TP to $\sim$\,15\% in the E-AGB
CO WD model, Table~\ref{hydro_mod}).

\begin{landscape}
\begin{deluxetable}{ccccccccccccc}
\tablecolumns{12} \tabletypesize{\footnotesize}
\tablecaption{Summary of the physical parameters of the four
  cases. These are the mass of the CO WD ($M_\mathrm{CO}$), He WD
  ($M_\mathrm{He}$), core ($m_\mathrm{Core}$), buffer zone
  ($m_\mathrm{buffer}$), SOF ($m_\mathrm{SOF}$) and envelope
  ($m_\mathrm{env}$). Columns 4 and 5 specify the progenitor AGB
  models used to construct the CO WD and the percentage of dredge-up
  from within the CO WD. Column 10 and 11 give the duration
  ($\tau_\mathrm{SOF}$) and the temperature
    ($T_\mathrm{SOF}/10^8\mathrm{K}$) of burning of the SOF, while 12
  and 13 indicate the duration the model spends in the RCB phase (in
  units of $10^5\mathrm{yr}$) and the colour representation of the
  cases in Fig.\,\ref{x_mod}. All masses $M$ and $m$ in $\mathrm{M_\odot}$.}
\tablewidth{0pt} \tablehead{ \colhead{CASE} &
  \colhead{$M_\mathrm{CO}$} & \colhead{$M_\mathrm{He}$} &
  \colhead{AGB} & \colhead{$f_{dre}$\,(\%)} &
  \colhead{$m_\mathrm{Core}$} & \colhead{$m_\mathrm{buffer}$} &
  \colhead{$m_\mathrm{SOF}$} & \colhead{$m_\mathrm{env}$} &
  \colhead{$\tau_\mathrm{SOF}$}& \colhead{$T_\mathrm{8,SOF}$} &
  \colhead{$t_\mathrm{5,RCB}$} & \colhead{colour}} \startdata
1 & 0.53 & 0.37 & E-AGB & 15.0 & 0.4505 & 0.0295 & 0.1000 & 0.32 & short &1.23& 1.02 & {\color{BurntOrange}orange}\\
2 & 0.53 & 0.37 & E-AGB & 15.0 & 0.4505 & 0.0295 & 0.1000 & 0.32 & long &1.23 &1.11 & {\color{Red}red}\\
3 & 0.53 & 0.37 & A-TP AGB & 5.4 & 0.5014 & 0.0107 & 0.0346 & 0.35 & short &1.23& 0.97 & {\color{Blue}blue}\\
4 & 0.60 & 0.30 & E-AGB & 15.0 & 0.5100 & 0.0141 & 0.1060 & 0.26 & long &2.42 & 2.75 & {\color{OliveGreen}green}\\
\enddata
\label{hydro_mod}
\end{deluxetable}
\end{landscape}

We consider two mass ratios from paper I, i.e.\ the $q=0.7$ case
($M_\mathrm{ CO}$ = 0.53$\,M_\mathrm{\odot}$ and $M_\mathrm{ He}$ =
0.37$\,M_\mathrm{\odot}$) with an SOF of $T=1.23\times10^{8}$\,K and
$\rho=5.02\times10^{4}$ \,$\mathrm{gcm^{-3}}$, and the $q=0.5$ case
($M_\mathrm{ CO}$ = 0.60$\,M_\mathrm{\odot}$ and $M_\mathrm{ He}$ =
0.30$\,M_\mathrm{\odot}$) with an SOF of $T=2.42\times10^{8}$\,K and
$\rho=3.16\times10^{4}$ \,$\mathrm{gcm^{-3}}$.  The buffer zone has a
temperature of approximately $5\times10^{7} \,\mathrm{K}$ and the same
density as the SOF, which is not hot enough for nucleosynthesis on the
dynamic time scale. Similarly, the core and envelope are too cold
for nuclear reactions during the merger.  The duration of the high-T
conditions in the SOF may extend beyond the time period covered by the
hydrodynamic simulations, at the end of which the SOF is still hot. In
addition, as pointed out in paper I, those simulations ignore any
feedback of nuclear energy production during the hydrodynamic phase,
which may affect the conditions and duration of the SOF. In order
to project the impact of different SOF assumptions, we consider a short
SOF with a duration of $8.2 \times 10^{5}\mathrm{s}$ and a long SOF
with a duration of $4.7 \times 10^{6}\mathrm{s}$, both of which are
within the viscous timescale determined by \citet{shen12}.

The four initial abundance profiles combine two CO-WD progenitor models, 
two mass ratios ($q=0.5$ and 0.7) with their respective SOF parameters
($T$, $\rho$ and the proportion of CO and He WDs in them), as well as
the short or long duration of SOF burning, as summarized in Table \ref{hydro_mod}.

\subsection{Stellar evolution and nucleosynthesis simulations}
For the post-merger evolution, we use version 3851 of the stellar evolution code
MESA \citep[][]{paxton11}. Initial models are created
from a homogeneous $0.90 M_\mathrm{\odot}$ (the total mass of the WDs
in the hydrodynamic simulations of paper I) zero-age main-sequence
model. This model is then relaxed to each of the four-zone post-merger
abundance profiles.
	
The star is evolved with a mass loss rate obtained by setting the
coefficient of Bl\"ocker's wind formula to $\eta = 0.05$, similar to
what is used in the NuGrid AGB simulations. From the options available
in MESA, we use the OPAL type I opacities for a solar abundance
distribution rather than the more appropriate type II opacities that
allow C and O enhancements from He-burning in the composition. This
choice may seem odd, since the irregular dust-ejections associated
with the unusual C dominated surface composition represent a defining
property of RCB stars. However, with our one-dimensional simulation
approach for the post-merger phase we can not hope in any case to
reproduce the irregular, and most likely aspherical mass ejection and
dust formation processes. In our stellar evolution simulations mass
loss is therefore only imposed via the Bl\"ocker formula with an
efficiency parameter choosen to obtain an order of magnitude
appropriate mass loss rate. Therefore, our choice of opacity will only
have minor effects on the stellar temperature during the RCB phase.

Indeed, individual test runs with type II opacities evolve to lower temperatures than
their type I counterparts, but the nucleosynthesis and
mixing evolution is in both cases the same. However, models with type
II opacities turn out to be much more difficult to converge, probably
reflecting some of the instabilities that do lead to the irregular
variability seen in RCB stars. Further,
the type II opacity runs did not show differences in the envelope
convection properties that would have altered our assumptions for the
adopted mixing profile (Section~\ref{sec:mixing}). Thus, the surface
abundance predictions remain the same in models with and without CO
enhanced opacities, and we choose to present here a homogeneous grid
of cases calculated with type I opacities.
	
Once the stellar evolution calculations are completed, the track is
post-processed for a complete nucleosynthesis analysis. We use the
NuGrid multi-zone nucleosynthesis code MPPNP \citep[][]{herwig08}
that processes each of the $\approx$\,1300 zones for each of the
$\approx$\,3000 time steps of the stellar evolution track with an
adaptive nuclear network that includes all stable as well as all relevant
unstable species (over 1000 isotopes) and their corresponding nuclear
reactions. The post-processing code also mixes species after each
time step, according to the mixing processes considered in the stellar evolution
simulations.

\subsection{The mixing model}
\label{sec:mixing}

\begin{figure}
\includegraphics[scale=0.75]{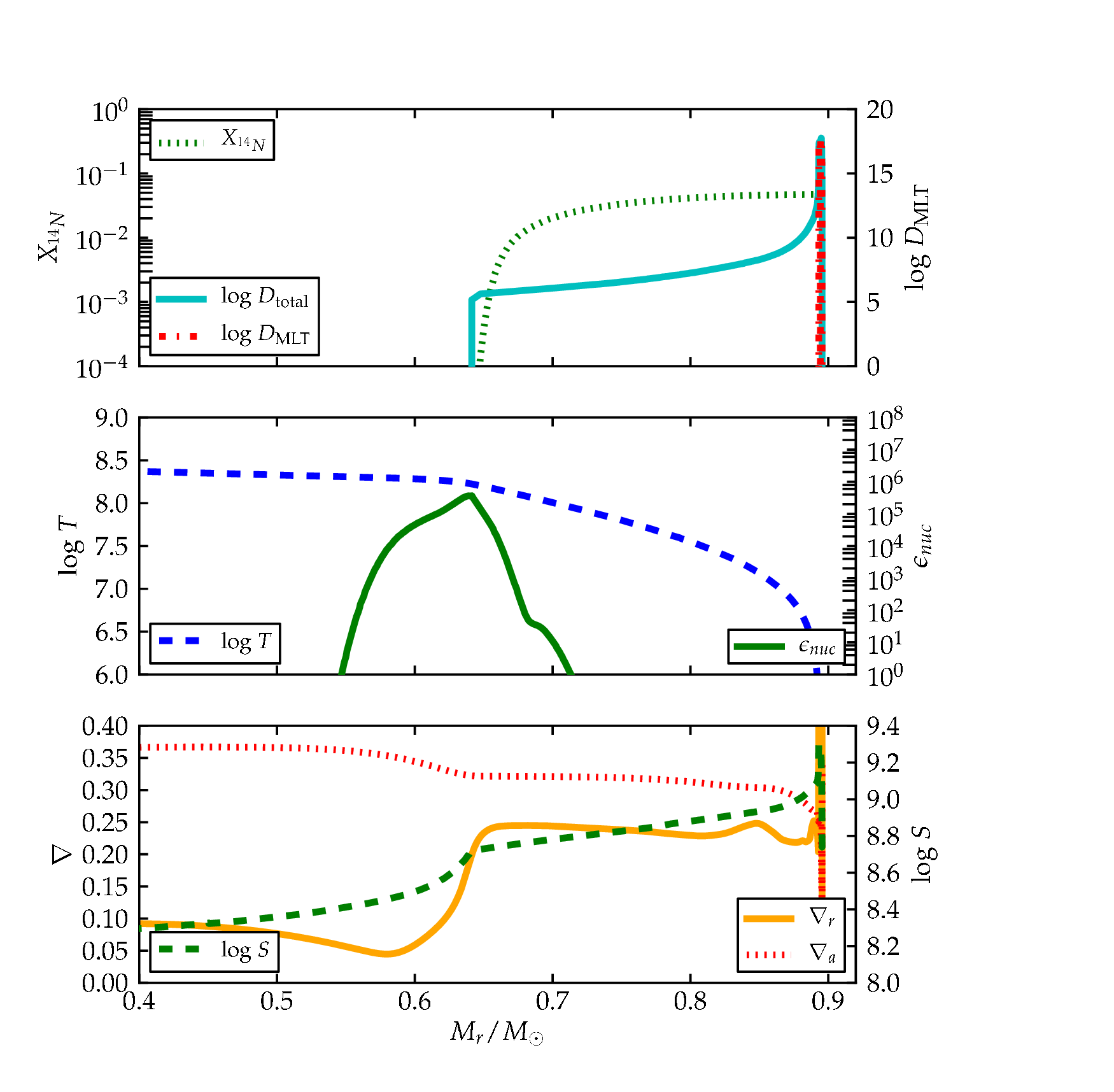}
\caption{The diffusion profile $D_\mathrm{total}$ at the timestep (iv) (Fig.\,\ref{part1}) for case 1 (black dot in Fig.\,\ref{HRD}), along with the MLT diffusion coefficient $D_\mathrm{MLT}$, in comparison with relevant physical parameters, such as the  $^{14}\mathrm{N}$ abundance ($X_\mathrm{^{14}N_{k}}$, panel 1), temperature (log $T$) and nuclear energy generation rate ($\epsilon_\mathrm{nuc}$) (panel 2), as well as the radiative ($\nabla_{r}$) and adiabatic ($\nabla_{a}$) temperature gradients and entropy (log $\,S$) (panel 3).}
\label{D_profile}
\end{figure}

\begin{figure}
\includegraphics[scale=0.75]{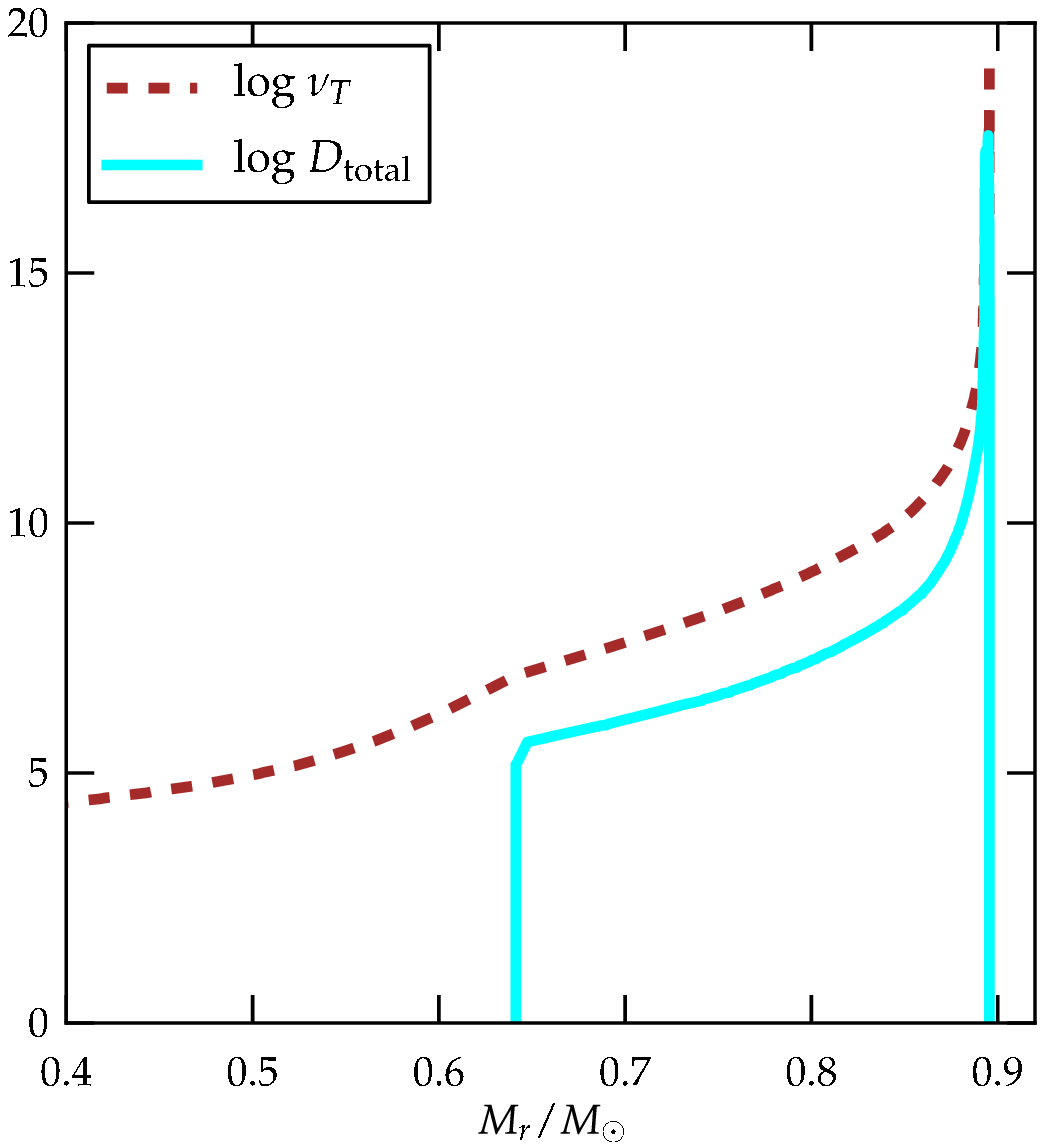}
\caption{The diffusion coefficient profile (log $D_\mathrm{total}$) at the timestep (iv) (Fig.\,\ref{part1}) for case 1 (black dot in Fig.\,\ref{HRD}) along with the thermal diffusivity (log $\nu_\mathrm{T}$).}
\label{D_vt}
\end{figure}

RCB stars show abundance signatures at their surfaces that originate
either in the SOF or in the nuclear processes during the post-merger
evolution. Mixing in the envelope reaching into the nuclear
processing region needs to bring such nuclear processed material to the
surface. Early on in our investigation it became clear that convection
alone would not be sufficient to provide this
essential mixing ingredient in the envelope.

The physical mechanisms of non-convective mixing processes that may
occur during or after a merger are quantitatively not understood.  In
any case, the merger remnant is rotating, and it is reasonable to
associate this rotation with mixing processes. \citet{shen12} estimate
that differential rotation in such an object would last for at least
$10^{4} - 10^{8}$\,seconds after the merging phase, after which it
would evolve towards solid-body rotation \citep{schwab12}. The exact
timescale for evolution towards solid-body rotation may be uncertain,
but some mixing processes, such as meridional circulation, can be driven by
solid body rotation as well.

Considering the significant uncertainties regarding both the realistic
rotation profile and the physics of rotation-driven mixing, we adopt an empirical
approach to additional mixing processes in this study.
This means that our primary goal is to identify a reasonable mixing law (a diffusion coefficient
profile) that is capable of reproducing the observed abundance patterns in RCB stars.

The diffusion coefficient in our mixing model ($D_\mathrm{total}$)
consists of convective mixing (${D_\mathrm{MLT}}$) predicted by
the mixing length theory (MLT) and an additional mixing
(${D_\mathrm{add}}$) component. The mixing is always restricted to the
region above $m_\mathrm{min}$ (\,=\,$m_\mathrm{Core}$ in each case)
(Eqs.~\ref{eq:mass_eq} and
Table~\ref{hydro_mod}). ${D_\mathrm{add}}$ is the Eulerian
representation of a constant mass diffusion coefficient
${D_\mathrm{Lag}}$ that drops sharply at the point where an entropy
barrier appears due to the peak of nuclear energy generation. The
location of this energy peak can be approximated as the
mass-coordinate corresponding to a specified fraction of the $^{14}$N abundance that
has still remained at this depth during He-burning, $k =
X(\mathrm{^{14}N_{k}})/X(\mathrm{^{14}N_\mathrm{max}})$, where
$X(\mathrm{^{14}N_{k}})$ is the mass fraction of $^{14}\mathrm{N}$ at
the mass coordinate $m_{k}$ where the mixing drops, and
$X_\mathrm{^{14}N_\mathrm{max}}$ is the maximum $^{14}\mathrm{N}$
mass fraction in the model at that time (Eqs.~\ref{eq:Dmix},
Fig.\,\ref{D_profile}).


The sharp drop of $D_\mathrm{total}$ at the mass-coordinate
$m_{k}$ is modeled by an exponential
decrease of the mixing efficiency.  This transition from the
extra-mixing zone to no-mixing is formally similar to how the
depth-dependent convective overshooting is implemented in the MESA code, and like in
the latter case, the width of this transition region is determined by the parameter $f$
in the exponent of Eq.~\ref{eq:f} (see the steep ramp in $D_\mathrm{total}$ in
Fig.\,\ref{D_profile}).

\begin{subequations}
\label{eq:Dmix}
\begin{align}
 D_\mathrm{total} &= 0  && m < m_\mathrm{min}  \\
 D_\mathrm{total} &= D_\mathrm{add}+D_\mathrm{MLT}  && m \ge
 m_\mathrm{min} \\ 
D_\mathrm{add}&= \frac{D_\mathrm{Lag}}{(4\pi r^{2}
  \rho)^{2}}\times \exp(\frac{-(T-T_\mathrm{k}))}{fT_\mathrm{k}})
&&m_\mathrm{min} \leq m \leq m_\mathrm{k} \label{eq:f}
\\
 D_\mathrm{add}&= \frac{D_\mathrm{Lag}}{(4\pi r^{2} \rho)^{2}} && m
 \geq m_\mathrm{k}, \label{eq:D4}
\end{align}
\end{subequations}
where $T_\mathrm{k}$ is the temperature at the mass coordinate
$m_{k}$. This mixing model is specified with the following parameters:
$D_\mathrm{Lag}= 4.5\times\mathrm{10^{51}}\,\mathrm{g^2/s}$, $k$=0.003
and $f$=0.05. The resulting Eulerian diffusion profile is shown in
Fig.\,\ref{D_profile}. In Fig.\,\ref{D_vt},
$\mathrm{D_\mathrm{total}}$ is compared with the thermal diffusivity
$\nu_\mathrm{T}$ for one of our considered cases. It is seen that
$\mathrm{D_\mathrm{total}}\ll\nu_\mathrm{T}$ everywhere inside the
star, which means that the additional mixing could be driven by a
secular instability rather than by a dynamical one. It is interesting
that the ratio between $\mathrm{D_\mathrm{total}}$ and
$\nu_\mathrm{T}$,
$\mathrm{D_\mathrm{total}}\sim10^{-2}\nu_\mathrm{T}$, is close to that
used by \citet{den08} to model extra mixing in low-mass red giants.

The effects of varying the free parameters in Eqs.\,\ref{eq:Dmix}
are briefly summarized here. When we increase $\mathrm{D_\mathrm{Lag}}$
by more than two orders of magnitude, the mixing becomes so fast that it causes an excessive
increase in $^{12}\mathrm{C}$ and $^{16}\mathrm{O}$ abundances at the surface and
a proportional decrease in $^{14}\mathrm{N}$ that disagree with
observations. There is also a risk for the star to become more
enhanced in oxygen than in carbon, the opposite of which is observed in
almost all RCBs. Making $\mathrm{D_\mathrm{Lag}}$ lower by a factor of
$10$ causes the mixing to become far too slow in dredging up
$^{12}\mathrm{C}$ and $^{16}\mathrm{O}$ to the surface. As the star
enters the RCB phase, it becomes more N-rich than C, which again is
not observed for any RCB star (except for one).

The surface chemical abundances of the star vary within an order of
magnitude when the value of the parameter $k$ is varied between
$10^{-3}$ and $5\times 10^{-2}$. Below the mass coordinate $m_\mathrm{k}$,
the diffusion profile drops very steeply. Allowing mixing to proceed
deeper into and through the He-burning shell would cause an excessive
dredge-up of $^{16}\mathrm{O}$. In addition, $^{19}\mathrm{F}$ and
$^{18}\mathrm{O}$ produced by partial He-burning would be destroyed in
this case. On the other hand, limiting the mixing to layers entirely
above the He-burning shell would prevent these two species from being
formed. Note that the location $m_\mathrm{k}$ of diminishing mixing
efficiency is physically motivated to coincide with the location of a
steepening positive entropy gradient associated with the He-burning
shell.

A small ramp appears in ${D_\mathrm{total}}$ just as it nears
$m_\mathrm{k}$. The beginning of the slope of the ramp and its
steepness are decided by the rather sensitive parameter $f$ in the
exponent function of $T$. The ramp feature affects the content of
$^{19}\mathrm{F}$, $^{18}\mathrm{O}$ and $^{16}\mathrm{O}$ at the
surface (more discussed in Section~\ref{sec:ramp}). The optimal
abundances in the star at a given value of ${D_\mathrm{Lag}}$ and $k$
are maintained when $f$ is chosen between $0.03$ and $0.06$.

Finally, we see indications from comparison with observed RCB star
abundance features that the mixing law should be time dependent. In
order to reproduce all the features, mixing must be strongly reduced
shortly before the star enters the RCB region of the HR diagram, as
will be discussed in Section~\ref{sec:c-n-o}.

\section{Results} 
\label{sec:results}
In this section we describe our results, starting with the
nucleosynthesis in the shell of fire in order to define the initial
conditions for the post-merger stellar evolution simulations, and
finally we describe the nucleosynthesis in these post-merger evolution
models. The original simulation output is available for further
analysis, see Section~\ref{sec:app_data}.

\subsection{The Shell of Fire (SOF)}
\label{sec:sof}
The formation and evolution of the shell of fire feature in the
dynamic merger simulations has been described in detail in paper I,
where we have also presented nucleosynthesis simulations. Here we just need
to briefly review and update those results with a focus on the
conditions for the production of n-capture elements, such as the s-process elements.  
\begin{figure}[t]
\includegraphics[width=0.9\textwidth]{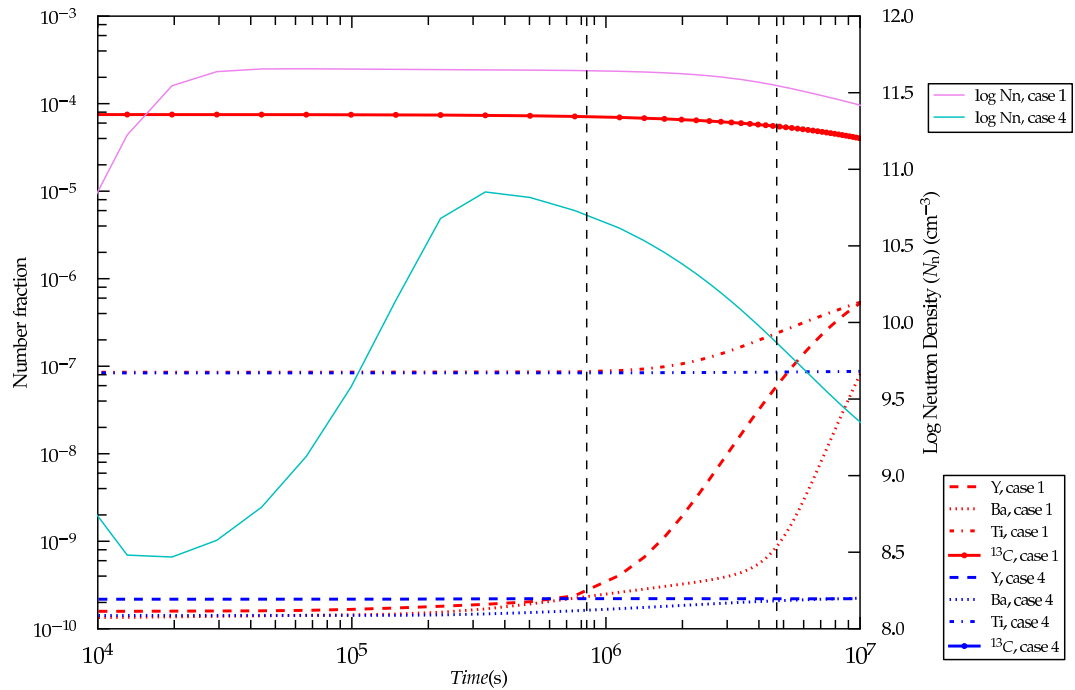}
\caption{The evolution of the abundances by number fraction of
$^{13}\mathrm{C}$(/100), Y, Ba and Ti during the burning of the SOF,
along with that of the neutron density, in the low-$T$ SOF used in case
1 (and 2, see Tab.\,\ref{hydro_mod}; case 3 assumes a different
progenitor) and the high-$T$ SOF used in case 4.  The vertical dashed
lines indicate the short-SOF ($8.2\times10^5\,s$) and long-SOF
($4.7\times10^6\,s$) durations. The abundance of $^{13}\mathrm{C}$ of
case 4 (blue) is beneath the lower limit of this plot.}
\label{sof_nden}
\end{figure}

The temperature in the SOF depends on the mass ratio $q$ of the merging
white dwarfs.  For the post-merger simulations presented here we
consider two SOF cases: case 4 is based on a high-$T$ ($2.42 \times
10^8 \mathrm{K}$, $q=0.5$) SOF while the other cases are based on a low-$T$ SOF
($1.23 \times 10^8 \mathrm{K}$, $q=0.7$).

In the low-$T$ SOF $^{13}\mathrm{C}$ forms early on at a level of
$\approx 10\%$ by mass and exceeds the abundance of $^{12}\mathrm{C}$
while $^{14}\mathrm{N}$ reaches $\approx 1\%$. In this situation
neutrons are steadily released via the
$^{13}\mathrm{C}\mathrm{(\alpha,n)}^{16}\mathrm{O}$ reaction with a
neutron density reaching a peak value of $\sim 3.16 \times 10^{11}
\mathrm{cm}^{-3}$ (Fig.\,\ref{sof_nden}). An interesting
neutron-recycling effect is at play here, in which protons released by
$^{14}\mathrm{N}(\mathrm{n},\mathrm{p}) ^{14}\mathrm{C}$ are mostly
creating another $^{13}\mathrm{C}$ and then again a neutron. This
explains why the $^{13}\mathrm{C}$ abundance shown in
Fig.\,\ref{sof_nden} decreases more slowly than expected considering just
the $\alpha$-capture reaction at that temperature.

In any case, sufficient neutrons are available to cause a steady
increase in the abundance of s-process elements along with nuclei such
as Ti. This production of n-capture elements becomes relevant only after
$\approx 10^{6}\mathrm{s}$ after the dynamic merger phase. Post-merger
evolutions have been considered for short and long SOF cases
(Table~\ref{hydro_mod}) and these differ in the amount of n-capture
elements originating in the SOF (see position of vertical lines in
Fig.\,\ref{sof_nden}). Only in the long-SOF case do s-process element
enhancements originate in the SOF, and provide abundances of n-caputure
elements that agree with RCB star observations.

In the high-$T$ SOF case $^{13}\mathrm{C}$ is destroyed by He-burning
within $10^{4}\,$seconds, and although $N_\mathrm{n}$ reaches a higher
peak value of $\sim 10^{14}\,\mathrm{cm}^{-3}$, it is immediately
reduced by neutron poisons, such as $^{14}\mathrm{N}$ and
$^{14}\mathrm{O}$ (Eq.\,11, Fig.\,15, paper I).  Later at
$10^{5}\,$seconds, the neutron density once again increases and
reaches a lower peak value of $\sim 6.31 \times
10^{10}\,\mathrm{cm}^{-3}$ through the neutron-source reaction
$^{17}\mathrm{O}\mathrm{(\alpha,n)}^{20}\mathrm{Ne}$. But these
neutrons are also not available for the production of s-process
elements due the presence of the $^{14}\mathrm{N}$ neutron poison. As
a result s-process elements and elements such as Ca and Ti are
unaffected in the high-$T$ SOF and remain close to their initial
values (Fig.\,\ref{sof_nden}).

\subsection{Stellar and nucleosynthesis evolution}
\subsubsection{The evolutionary tracks}
\label{sec:nuc}
\begin{figure}
\includegraphics[scale=0.80]{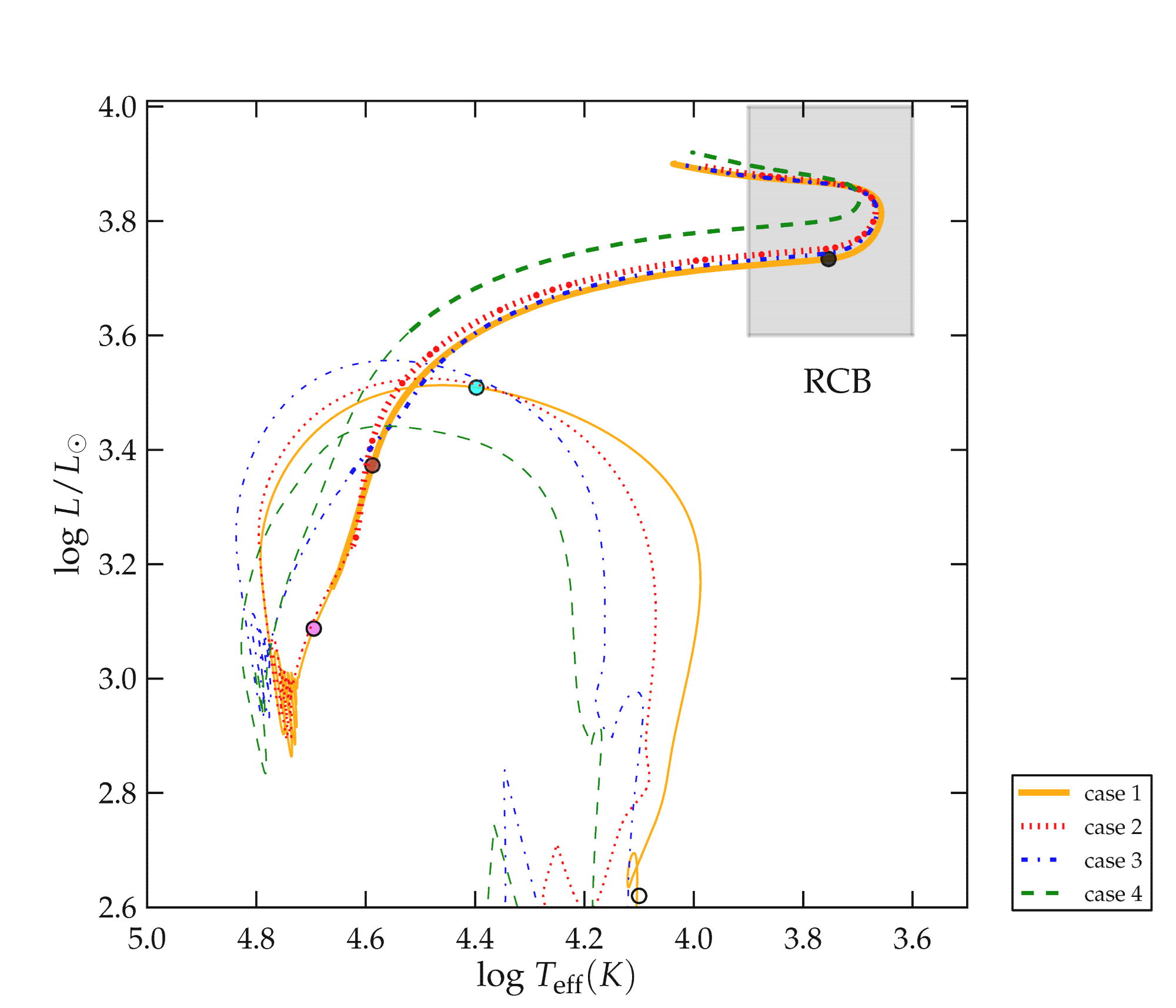}
\caption{The HR diagram of the four considered cases. The thinner portions of the lines indicate the evolution when 
the star burns H in a shell and the grey box shows the area of the HR diagram in which RCBs are observed to lie. 
The dots correspond to timesteps in the evolution:   (i) - 0\,years, empty black circle, 
(ii) - $4.2 \times \mathrm{10^{5}}$\,years, {\color{Cyan}cyan} dot, (iii) - $6.7 \times \mathrm{10^{5}}$\,years, 
{\color{VioletRed}violet} dot, (iv) - $1.3 \times \mathrm{10^{6}}$\,years, {\color{Brown}brown} dot, and 
(v) - $7.2 \times \mathrm{10^{6}}$\,years, black dot.}
\label{HRD}
\end{figure}

The evolutionary tracks of the four cases start at low luminosity in
the HR diagram and are similar to each other (Fig.\,\ref{HRD}). The
thin lines in the lower portion of the plot correspond to the initial
H-shell burning period and are not expected to reproduce all details of
the ongoing thermal readjustment of the post-merger star. Fortunately,
the details of this part of the evolution are not important for the
RCB star surface abundance predictions. However, the
thicker lines indicate when He-shell burning takes over, and the star
is now in hydrostatic and thermal equilibrium and evolves on the
nuclear timescale of the He-burning shell. Observed RCB stars can be
found in the range $3000\,\mathrm{K}\leq T_\mathrm{eff}\leq 8000\,\mathrm{K}$
and $3.5\leq\log L/L_\mathrm{\odot}\leq 4.0$ \citep{clayton96,pandey08}.
 
The models spend between $0.1$ (case 3) and $2.75 \times 10^5\,
\mathrm{yr}$ (case 4) in the RCB phase. This age is obtained when the mass-loss rate is 
$\sim 10^{-7}\,M_\mathrm{\odot}/\mathrm{yr}$. Increasing the mass-loss rate decreases 
the time spent in the RCB phase. A mass-loss rate of 
$\sim 10^{-5}\,M_\mathrm{\odot}/\mathrm{yr}$ would reduce the period of the
RCB phase of the model to $\sim 10^4$\,years, which is the expected
period of the RCB star phase \citep{clayton12}.

\begin{figure}
\includegraphics[scale=0.50]{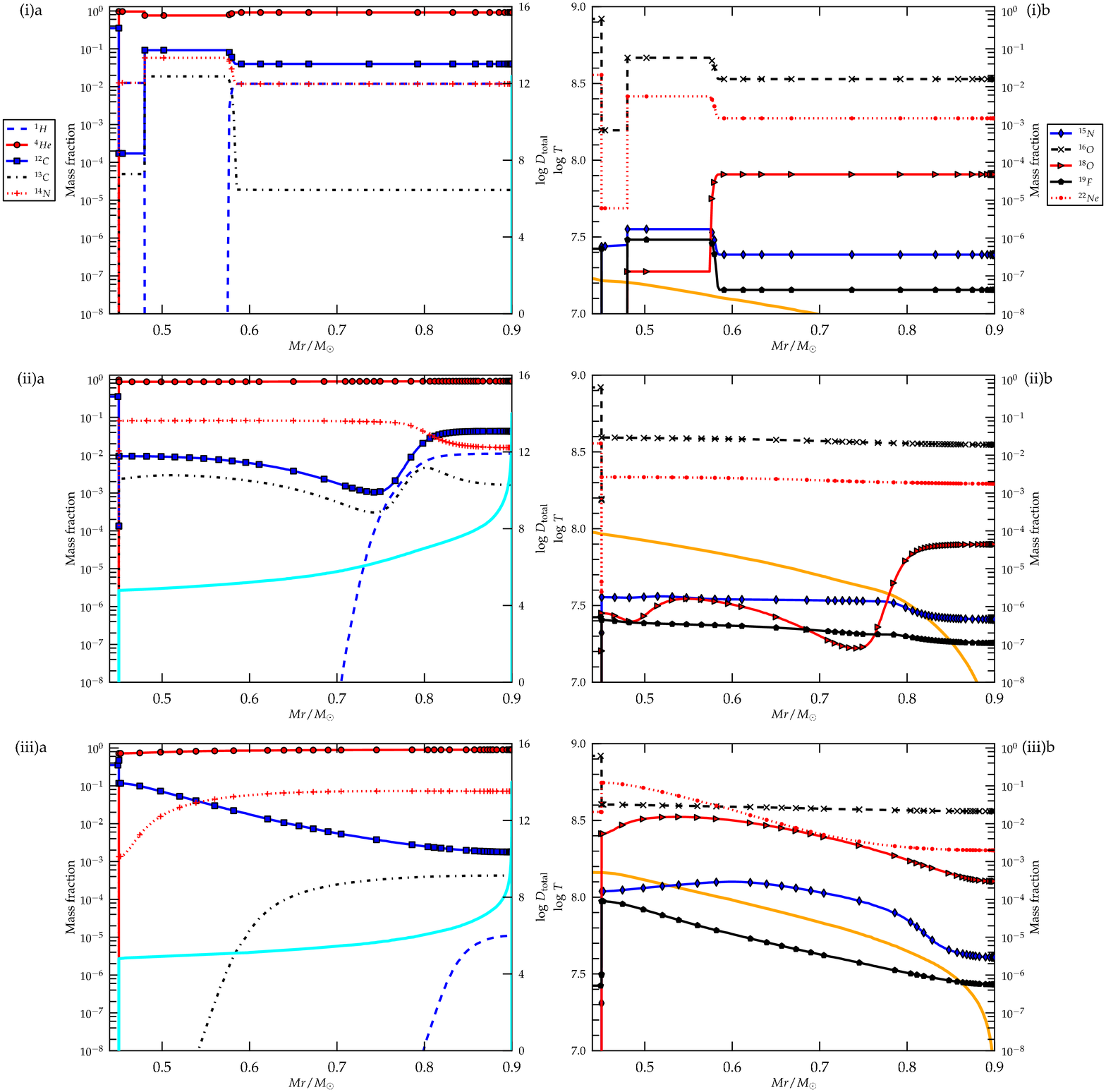}
\caption{The abundance distributions for nuclear species between $^{1}
  \mathrm{H}$ and $^{22}\mathrm{Ne}$ at three timesteps of evolution
  for case 1.  The panels correspond to the dots in
  Fig.\,\ref{HRD}. The line of every nuclear species connects the
  abundance of that species at every $20^\mathrm{th}$ zone in the
  model. The panels in the left column have log $D_\mathrm{total}$
  (solid cyan line) plotted in them and those in the right column have
  log $T$ (solid orange line).}
\label{part1}
\end{figure}
\begin{figure}
\includegraphics[width=1.0\textwidth]{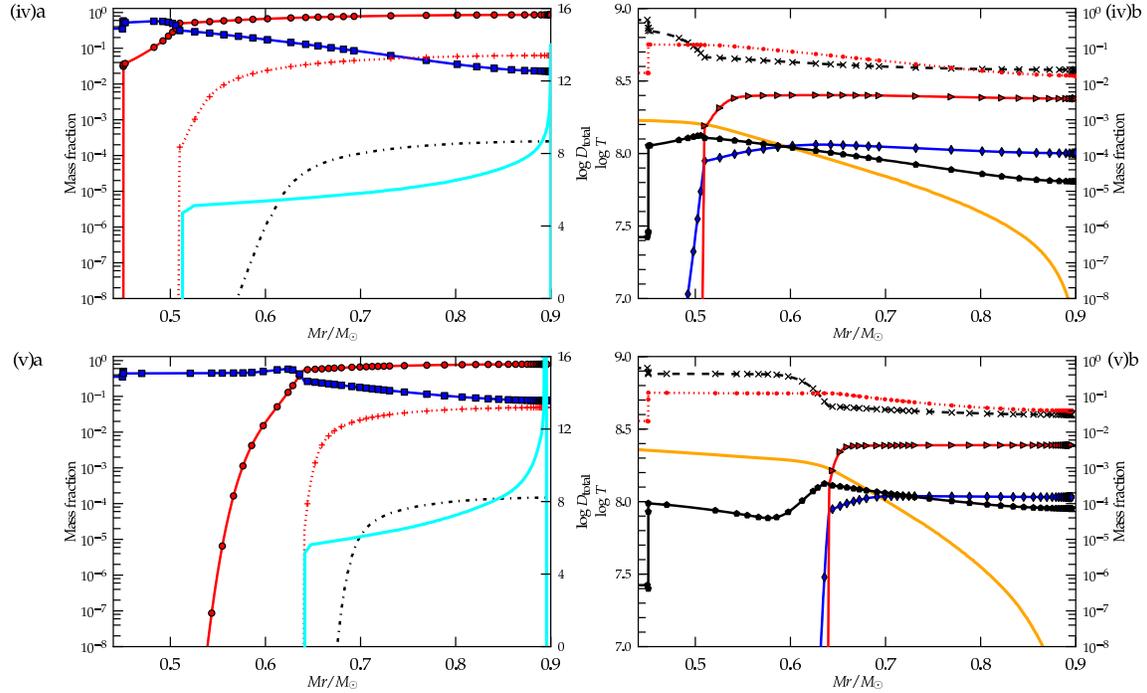}
\caption{The abundance distributions for nuclear species between $^{1}
  \mathrm{H}$ and $^{22}\mathrm{Ne}$ at two timesteps of evolution for
  case 1. The panels correspond to the dots in Fig.\,\ref{HRD}.  The
  line of every nuclear species connects the abundance of that species
  at every $20^\mathrm{th}$ zone in the model.  The panels in the left
  column have log $D_\mathrm{total}$ (solid cyan line) plotted in them
  and those in the right column have log $T$ (solid orange line).}
\label{part2}
\end{figure}

\subsubsection{Nucleosynthesis and mixing: origin of 
  $^{18}\mathrm{O}$ and $^{19}\mathrm{F}$ }
\label{sec:ramp}

We discuss the different phases of burning and associated
nucleosynthesis processes that occur in the post-merger evolution of our
models by focusing on case 1 (Table~\ref{hydro_mod}).  Dots along the
HRD track in Fig.\,\ref{HRD} correspond to the profiles shown in
Fig.\,\ref{part1}.  Figs.\,\ref{part1}(i)a and b show the initial
abundance profiles of additional isotopes (compared to
Fig.\,\ref{ini_profile}) for case 1. The second set of panels
(corresponding to the cyan dot in Fig.\,\ref{HRD}) shows CNO-cycle
burning of the hydrogen contributed to the merged CO+He WD system from
the H-rich envelope of the original He WD, which was uniformly
distributed throughout the He WD in our construction of the initial
model (cf.\ Section~\ref{sec:initial_comp}). The maximum temperature
during this phase is $\sim 10^{7}$\,K, and the H-shell leaves
$^{14}\mathrm{N}$ (via $^{13}\mathrm{C(p,\gamma)}^{14}\mathrm{N}$) and
$^{13}\mathrm{C}$ (via
$^{12}\mathrm{C(p,\gamma)}^{13}\mathrm{N(\beta^+)}^{13}\mathrm{C}$)
behind. This phase results in an overall increase in the abundances of
$^{13}\mathrm{C}$ and $^{14}\mathrm{N}$ in the envelope.

Eventually, the temperature at the core-envelope interface increases
further to $\sim 10^{8}\,\mathrm{K}$, and beginning at
$m_\mathrm{min}$ (cf.\ Section~\ref{sec:mixing}), He-shell burning
starts (violet dot in Fig.\,\ref{HRD}). By this time, the H-shell has
almost completely burnt outwards to retreat to a thin layer close to
the surface (Fig.\,\ref{part1}(ii)a). The first neutron burst, via
$\alpha$-capture on $^{13}\mathrm{C}$ is triggered at this time, and
consequently the $^{13}\mathrm{C}$-shell moves outwards as it gets
burnt to $^{16}\mathrm{O}$.

In the time between panels (ii) and (iii) of Fig.\,\ref{part1}
(violet and brown dots in Fig.\,\ref{HRD}), partial He-burning of
$^{14}\mathrm{N}$ raises the $^{18}\mathrm{O}$ to its peak value, where
$T\sim  1.20\times10^{8}$\,K, while some of this $^{18}\mathrm{O}$
begins to get destroyed to $^{22}\mathrm{Ne}$ by further
$\alpha$-capture closer to $m_\mathrm{min}$. 

As He-burning starts, the $^{13}\mathrm{C}(\alpha,\mathrm{n})
^{16}\mathrm{O}$ reaction releases neutrons at a low density of
$N_\mathrm{n}\approx \mathrm{10^{5}}$\,cm$^{-3}$. This is too little
to produce s-process elements, in particular in view of the large
$^{14}\mathrm{N}$ abundance that acts as a neutron poison for the s
process. However, the neutron poison reaction enables the production
of $^{19}\mathrm{F}$.  The abundances of $^{15}\mathrm{N}$ and
$^{19}\mathrm{F}$ increase via the reaction chain
$^{14}\mathrm{N\mathrm(n,p)}^{14}\mathrm{C\mathrm(p,\gamma)}^{15}\mathrm{N(\alpha,\gamma)}^{19}\mathrm{F}$. Panel
(iii) shows how this reaction chain is able to produce
$^{15}\mathrm{N}$ and $^{19}\mathrm{F}$ efficiently in this
\emph{simultaneous slow mixing and burning environment}. The challenge
for any scenario for the simultaneous production of $^{19}\mathrm{F}$
and $^{18}\mathrm{O}$ is the relative efficiency of the
$\alpha$-capture, which is increasing in the order $^{13}\mathrm{C}$,
$^{18}\mathrm{O}$, $^{15}\mathrm{N}$ and finally
$^{19}\mathrm{F}$. $^{15}\mathrm{N}$ appears in the more external
layers, where $^{13}\mathrm{C}$ releases neutrons. However,
$^{19}\mathrm{F}$ can only form from $^{15}\mathrm{N}$ in deep enough
layers, where typically $^{18}\mathrm{O}$ would already be
destroyed. Therefore, mixing must be fast enough to bring
$^{15}\mathrm{N}$ below the region where $^{18}\mathrm{O}$ can form,
but slow enough to not transport all that $^{18}\mathrm{O}$ down where
it would react with $\alpha$ particles into $^{22}\mathrm{Ne}$.

This balance between mixing processes and nucleosynthesis is also
realized in the ramp region (Eq.\,\ref{eq:f}) of our adopted mixing
profile. Such a continuous transition from the envelope mixing regime
to the non-mixed core is a physically reasonable assumption, and it
has to be emphasized that without such a ramp transition our models
can not reproduce the observed abundances of F as well as they do. It is
this ramp transition region that is responsible for much of the
continuous increase in $^{19}\mathrm{F}$ during the later part of the
evolution of our model stars to the RCB star regime (panel (iv) and (v) in
Fig.\,\ref{part2}, black dot in Fig.\,\ref{HRD}).

If we assume that the diffusion coefficient drops more steeply at
$m_\mathrm{k}$ (via the parameter $f$ in Eq.\,\ref{eq:f}), then less
$^{19}\mathrm{F}$ is produced. For a larger $f$ parameter, the mixing
profile extends deeper and the surface amount of $^{18}\mathrm{O}$
decreases, while that of $^{16}\mathrm{O}$ increases.  Therefore, the
combined abundance evolution of $^{18}\mathrm{O}$ and
$^{19}\mathrm{F}$ provide strong limits on the width of the transition
region between the mixed envelope and unmixed core at and below
the He-burning shell. With our choice of $f$ the models are allowed to
produce the maximum amount of $^{19}\mathrm{F}$ without destroying too
much $^{18}\mathrm{O}$ and increasing the $^{16}\mathrm{O}$ at the
surface. The ramp-transition feature is thus a sensitive, integral and
unavoidable part of the mixing profile.

In panel (iv), Fig.\,\ref{part2}, $T$ has risen at $m_\mathrm{min}$ to
$\sim 3.16 \times 10^{8}$\,K, and the triple-$\alpha$ chain has been
fully activated causing $^{12}\mathrm{C}$ to increase and a further
$\alpha$-capture causing $^{16}\mathrm{O}$ to increase
(Fig.\,\ref{part1}(iv)a). Below $m_{k}$, at which the mixing profile
drops, complete He-burning of $^{18}\mathrm{O}$ and $^{19}\mathrm{F}$ to
$^{22}\mathrm{Ne}$ takes place. By the time the star enters the RCB
phase in the HR diagram, it has a strong He-shell burning (panel (v) of
Fig.\,\ref{part2}, black dot in Fig.\,\ref{HRD}). Between panels (iv)
and (v), a second weaker neutron-source, that of $^{22}\mathrm{Ne}$, is
initiated through $^{22}\mathrm{Ne(\alpha,n)}^{25}\mathrm{Mg}$. The
neutron densities ($N_\mathrm{n}$) during the post-merger evolution
range between $10^4$ and $10^6\,\,\mathrm{cm}^{-3}$.

The decrease in mixing efficiency follows the He-burning shell by
design (cf. Section~\ref{sec:mixing}, Eqs.\,\ref{eq:Dmix}). This
design prevents the material that is processed by strong He-shell
burning, such as $^{12}\mathrm{C}$ and, particularly, $^{16}\mathrm{O}$,
from reaching the surface (panels (iv)a and (v)a, Fig.\,\ref{part2}),
where it would cause surface abundance changes that are not
observed. The design also prevents $^{18}\mathrm{O}$ and
$^{19}\mathrm{F}$ isotopes from above $m_\mathrm{k}$ to be mixed to
the higher temperatures in the interior, where they would be easily
destroyed. Thus, the formulation of the mixing profile avoids an excess
pollution of the surface by $^{16}\mathrm{O}$ and preserves the
$^{18}\mathrm{O}$ and $^{19}\mathrm{F}$ abundances in the envelope. As explained
in Section~\ref{sec:mixing}, the fast decrease in mixing efficiency in
our mixing model is motivated by the positive entropy gradient induced
by the He-burning shell.

 
The results of similar simulations for the other three cases from Table~\ref{hydro_mod} are 
shown in Figs.\,\ref{cno}\,--\,\ref{x_rcb} and they are compared with
observations in Section~\ref{sec:surf-abun}.

\subsubsection{s process in the post-merger evolution}
\label{sec:sproc-pmerg}
\begin{figure}[t]
\includegraphics[width=1\textwidth]{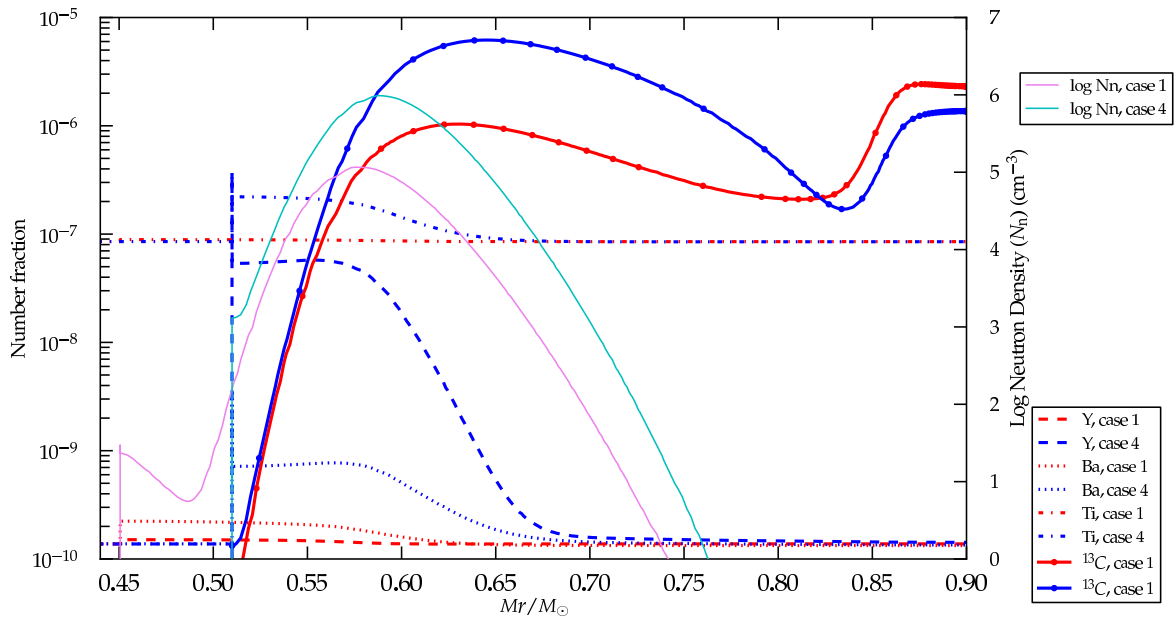}
\caption{The abundances by number fraction of $^{13}\mathrm{C}/100$,
  Y, Ba and Ti during the post-merger evolution of cases 1 and 4,
  along with neutron densities. The snapshots of time for these
  profiles are at $3.1 \times 10^5\mathrm{yr}$ (case 4) and $4.5 \times
  10^5\mathrm{yr}$ (case 1), between the cyan and the violet dots in
  Fig.\,\ref{HRD}.}
\label{post-merger_nden}
\end{figure}
As seen in the previous section n-capture nucleosynthesis plays an
important role in producing the observed F abundance. In case 4
($q=0.5$, high-$T$ SOF, no SOF s-process production, c.f.\
Section~\ref{sec:sof}) the neutron exposure in the post-merger evolution
is sufficient to produce s-process elements. In this low-$q$ case we
have deduced a higher dredge-up mass from the dynamic merger simulation
which implies a higher initial abundance of $^{12}\mathrm{C}$. The lower
He-WD mass of the low-$q$ case implies a larger WD envelope mass and
therefore a larger proton abundance in the post-merger envelope. Taken
together the low-$q$ post-merger evolution features a larger amount of
the n-source $^{13}\mathrm{C}$. This leads to a higher neutron density
in case 4 ($N_\mathrm{n}\approx10^{6}\,\mathrm{cm}^{-3}$) compared to
cases 1\,--\,3 ($N_\mathrm{n}\approx10^{4}-10^{5}\,\mathrm{cm}^{-3}$)
(Fig.\,\ref{post-merger_nden}). Thus, in the post-merger evolution of
case 4 the abundances of s-process elements are increased by up to $\sim
1 \mathrm{dex}$ (e.g.\ in the case of Y) from their initial values. In
cases 1\,--\,3 the post-merger evolution n-capture nucleosynthesis is
significantly less efficient.

\subsection{Comparison of the surface abundances in our models with
  observations} RCB stars are classified into two categories based on
their values of [Fe] and Si/Fe and S/Fe ratios \citep{lambert94}. RCB
minority stars (plotted as black squares, Fig.\,\ref{cno}) are
identified as having lower [Fe] and higher Si/Fe and S/Fe ratios than
the majority (plotted as black stars, Fig.\,\ref{cno}). We adopt the
data of measured elemental abundances of RCBs summarized by
\citet{jeffery11} and use the solar scaled log representation of
abundances, i.e.\ [X]= $\epsilon_{i}- \epsilon_{i_\mathrm{\odot}}$
where $\epsilon_{i}$ $\equiv$ log $n_{i} + C$, $n_{i}$ being the
relative number abundance of species $i$. This data summary does not
include the individual error bars on each data point, but in general
the errors are quoted to be $0.2 - 0.3 \mathrm{dex}$.

\subsubsection{C\,$>$\,N\,$>$\,O : constraints on the mixing}
\label{sec:c-n-o}
\begin{figure}
\includegraphics[scale=0.78]{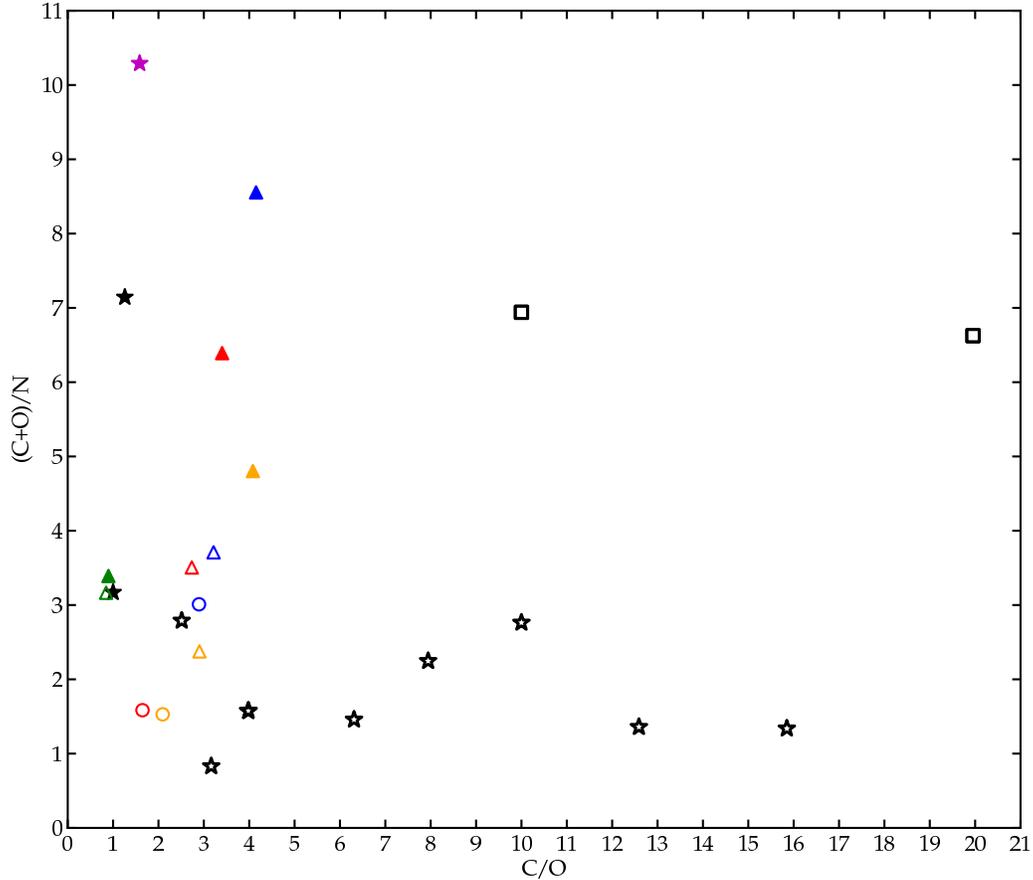}
\caption{The (C+O)/N ratio of the four computed cases (case 1
({\color{BurntOrange}orange} symbols), case 2 ({\color{Red}red}
symbols), case 3 ({\color{Blue}blue} symbols) and case 4
({\color{OliveGreen}green} symbols)) and in RCBs (minority - black
squares, majority - black stars) against their C/O ratio. Two minority
stars have been omitted due to their C/O ratio being out of the range of
the plot. R CrB is indicated by the magenta star in the plot. Plotted
for every case is the surface value of these ratios when the model is
within the RCB range (empty circles), when it enters the RCB phase in
the HR diagram (Fig.\,\ref{HRD}) (empty triangles) and when it leaves
this phase (filled triangles). All observed RCBs that have N$<$O in
their surface are indicated by filled symbols.}
\label{cno}
\end{figure}

The abundance of C+O traces He-burning, while the C/O ratio signals
the relative importance of the triple-$\alpha$ to
$^{12}\mathrm{C(\alpha,\gamma)}^{16}\mathrm{O}$ reactions which is
sensitive to temperature. The elemental abundance of N is the
result of CNO processing. Therefore, the (C+O)/N ratio indicates the
relative importance of He- to H-burning processes in determining the
observed abundance distribution in RCB stars. Their distribution
in the (C+O)/N vs. C/O plane shows that the majority lie
in a particular range of (C+O)/N ratio, of 1\,--\,3 and are more spread in
their C/O ratio, with most RCBs having a value of 1\,--\,10
(Fig.\,\ref{cno}). We find that on keeping the amount of mixing
constant at the initial value of $D_\mathrm{Lag}$ in our models, the
$^{14}\mathrm{N}$ in the envelope reduces continuously by being burnt
to $^{18}\mathrm{O}$ and this causes an overall decrease in the mass
fraction of nitrogen at the surface. On continuing the evolution of
the model at the same rate of mixing up to the end of the RCB phase,
the (C+O)/N ratio increases at least by up to a factor of 3 higher
than its value in most RCB stars (except for 3 stars in
Fig.\,\ref{cno} but these have N$>$O). If the rate of mixing is
lowered and restricted to smaller depths from the surface once the
model attains the (C+O)/N and C/O ratios of most RCBs (the empty
circles in Fig.\,\ref{cno}), the $^{14}\mathrm{N}$ on the surface is
preserved through the RCB phase of the star. Thus in order to attain
the distribution of C, N and O of RCBs, our models strongly indicate
that the magnitude of mixing and its reach from the surface must
decrease with time. \citet{schwab12} find that continuous
transportation of angular momentum from the merged system causes
differential rotation to reduce in the envelope. Although the
timescale over which they see this happens in their simulations is
shorter than what our models indicate, the idea that such a reduction
in mixing driven by rotation can happen is very feasible.

The difference in the surface abundances between restricting mixing at
the time when the model lies in the observed domain of (C+O)/N and C/O
and on continuing mixing as the star enters the RCB phase, is not
significant for elements between Na and La. However a considerable
difference of the order of $0.5 \mathrm{dex}$ does arise in the C, N, O and F
abundances as seen in in Fig.\,\ref{x_mod}. Case 4 does not at any
point of time in its evolution have a model with C$>$N$>$O on its
surface (hence the absence of circles in Fig.\,\ref{x_mod}) and its
O-content is almost the same as the C-content in its surface.

\subsubsection{Elemental surface abundances}
\label{sec:surf-abun}
\begin{figure}
\begin{centering}
\includegraphics[scale=0.7]{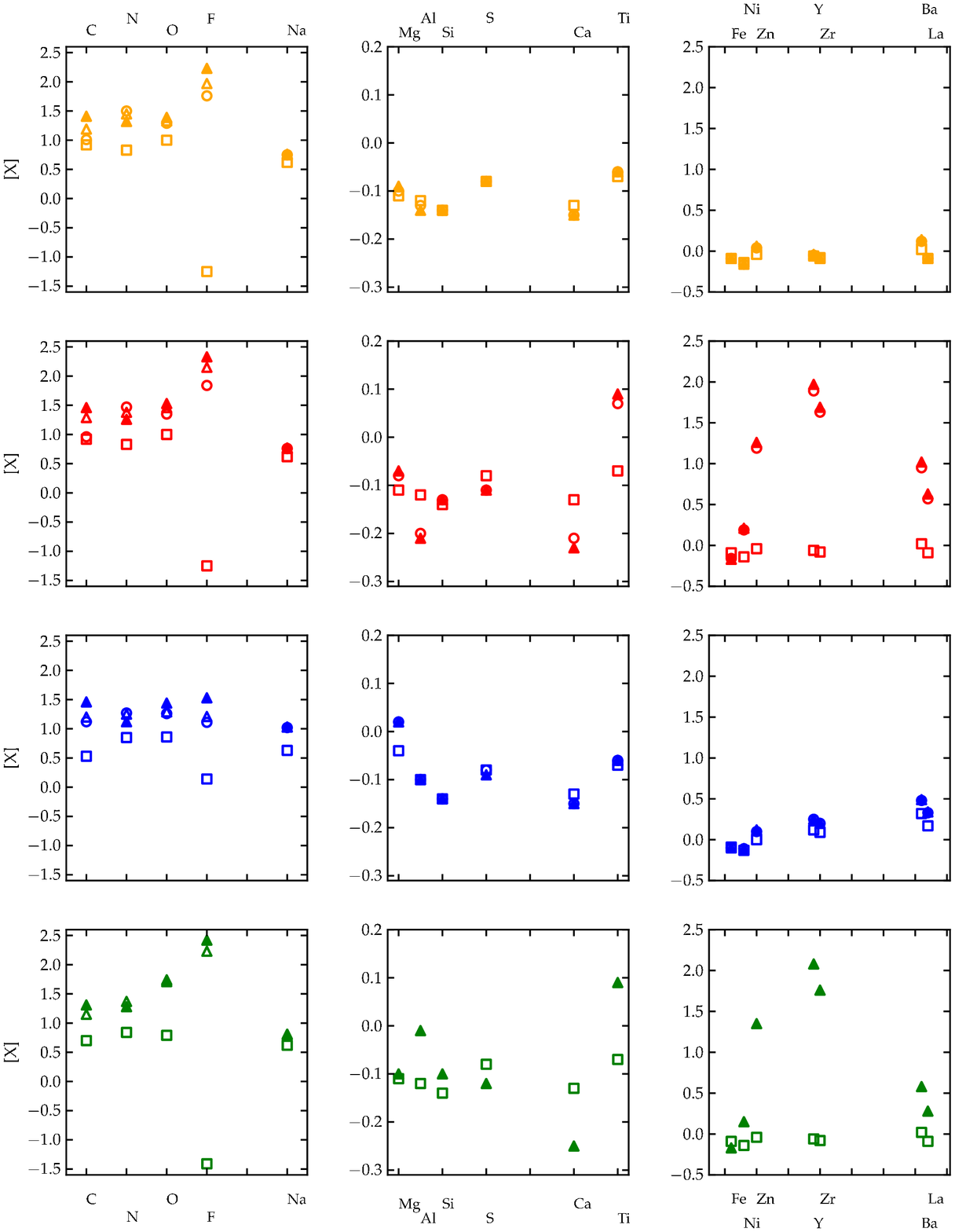}
\end{centering}
\caption{Surface [X] abundances of {\color{BurntOrange}case 1},
  {\color{Red}case 2}, {\color{Blue}case 4} and
  {\color{OliveGreen}case 4}. The empty squares stand for the initial
  envelope abundance while the meaning of the rest of the symbols are
  as explained in Fig.\,\ref{cno}. For species between Mg to La, the
  difference in abundances between the models that enter and exit the
  RCB phase is $< 0.2 \mathrm{dex}$, hence only the average abundances
  of these species in this phase are plotted (filled triangles).}
\label{x_mod}
\end{figure}

\begin{figure}
\begin{centering}
\includegraphics[width=1.05\textwidth]{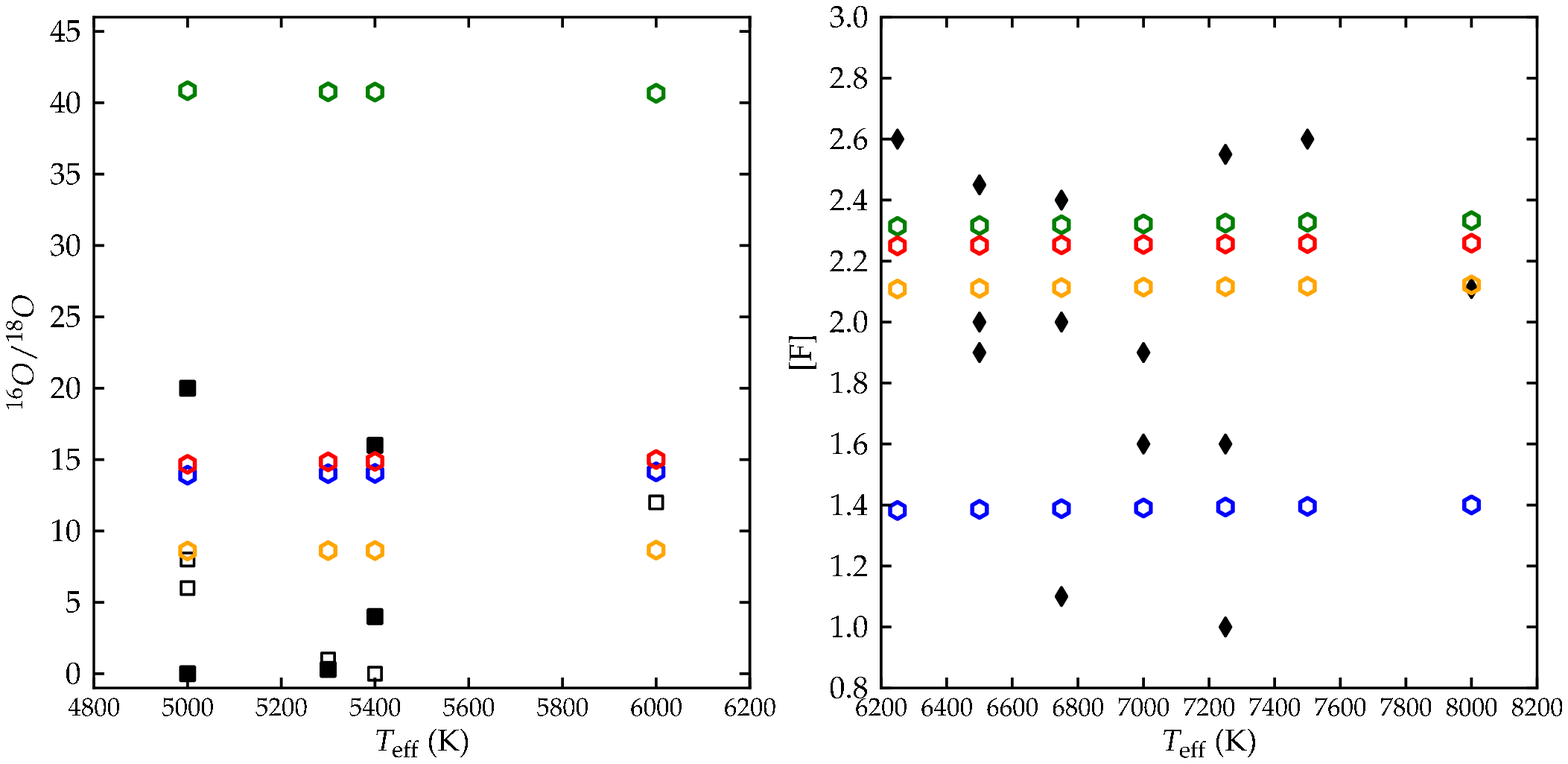}
\end{centering}
\caption{Surface $^{16}\mathrm{O} / ^{18}\mathrm{O}$ ratios (squares
  are the observed data; empty squares from \citet{clayton07} and
  filled ones from \citet{garcia10}) and [F] (black diamonds are the
  observed values from \citet{pandey08}) against $T_\mathrm{eff}$ for
  the four cases (colours are the same as in Fig.\,\ref{x_mod}). The
  model abundances of the nuclei plotted here (hexagons) are averaged
  between the timesteps at which the star enters and leaves the RCB
  phase (Fig.\,\ref{HRD}).}
\label{o_f}
\end{figure}

\begin{figure}
\includegraphics[width=1.05\textwidth]{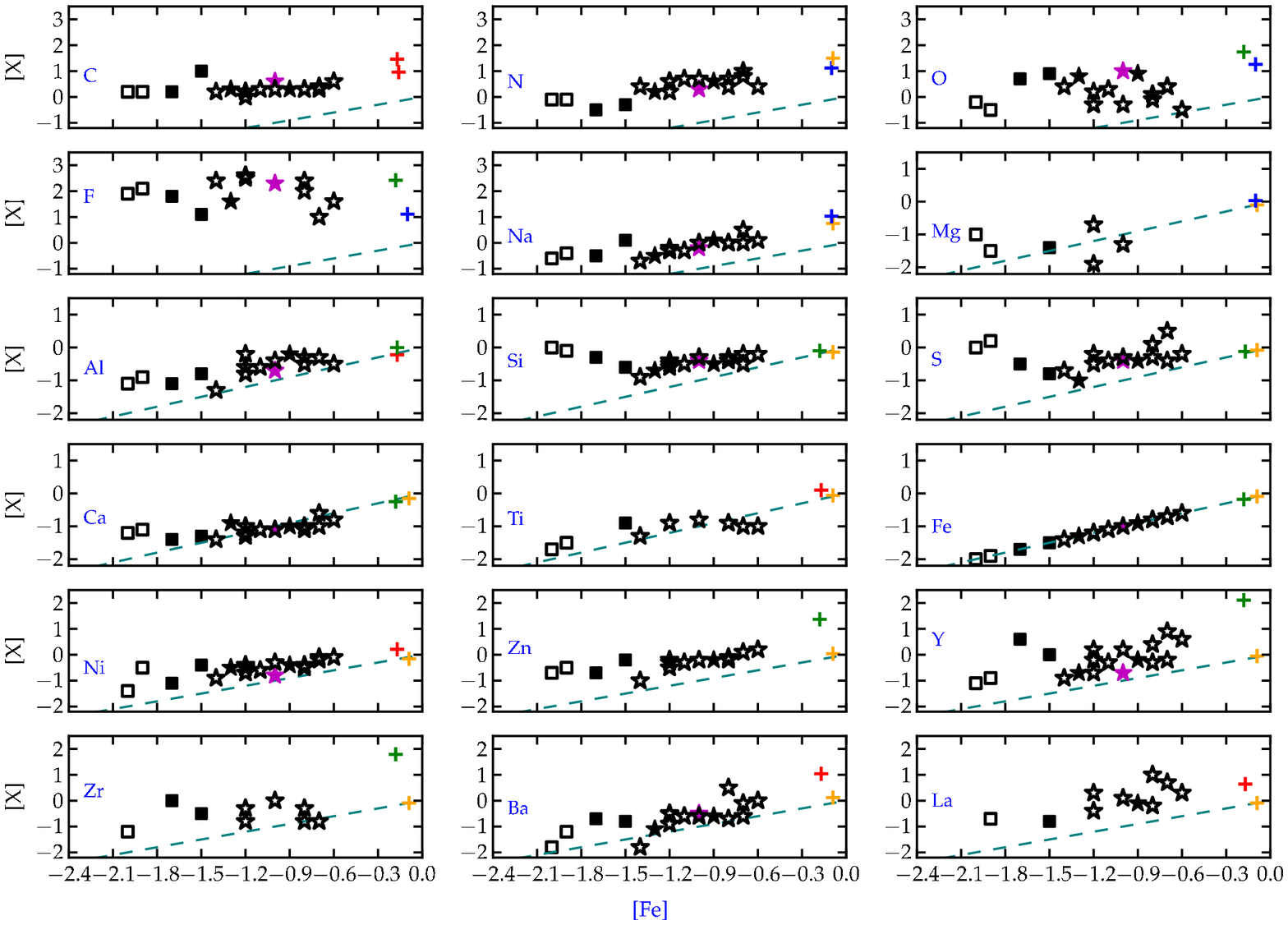}
\caption{The elemental [X] values in the observed RCBs and their range from our cases. The crosses represent the maximum and minimum values of the given element from the four cases (colors according to Fig.\,\ref{x_mod}) after the star becomes an RCB star. The dashed lines across each panel show the expected abundances by scaling the solar composition with [Fe].}
\label{x_rcb}
\end{figure}

In this section we provide an element by element discussion based on
the following diagrams. Fig.\,\ref{x_mod} provides a summary for
the surface abundance evolution through the RCB phase of all four
cases. Fig.\,\ref{o_f} compares the O-isotopic ratio and F abundance
of models with observations.
The [X] values of the observed nuclear species in majority and
minority RCBs are shown in Fig.\,\ref{x_rcb}, along with the lowest
and highest values of each element from the four cases, between the
time when star attains the observed C/N and N/O ratios in RCBs and
when it exits the RCB phase in Fig.\,\ref{HRD}.

As RCBs are associated with the old bulge population (intermediate Pop II, Sec.\,\ref{sec:intro}), we assume that their range of Fe abundances reflects their initial metallicity.  Our models were constructed with solar metallicity, which allowed us to take advantage of our detailed nucleosynthesis and stellar evolution progenitor models, currently available only at this metallicity. Fortunately, this does not affect the relevance of our models for comparison with observations in any significant way. The nuclear species in our models can be divided into primary and secondary types, depending on the nature of the processes that affect their abundances.  With this difference in mind our models can be compared with observations of RCB stars for a range of metallicities in the following way.

The primary elements are formed entirely from the burning of H and He and so, are independent of the initial [Fe] abundance of the star. Therefore, in Fig.\,\ref{x_rcb}, one must compare only between the [X] values of these elements between the models and the observations. 

The abundances of secondary elements in the model depend on the
proportion of their seed nuclei which were formed in previous
generations of stars. These include species between Na and Ti and the
s-process elements. Therefore, in Fig.\,\ref{x_rcb}, one must compare
the offset of the [X] value of the element with the solar scaled
composition shown as a dashed green line.

\paragraph{H and Li:}
 The high amount of hydrogen depletion reported for all RCBs, is a natural outcome of our models. But Li being a fragile element is easily destroyed by H- and He-burning during the merging phase of the WDs itself and consequently in our models too. Therefore we do not include H and Li in the discussion of surface abundance results. 

\subsubsubsection{Primary elements}
\paragraph{C, N, O and F:}
One of our main objectives in this work was to find the 
high enhancements in F and the low $^{16}\mathrm{O} / ^{18}\mathrm{O}$
ratios of RCBs. Fig.\,\ref{o_f} shows the good agreement between our
models and the observed range of these species. Three of the cases
have $^{16}\mathrm{O}$/$^{18}\mathrm{O}$ ratios between 6 and 15 that are
within the measured range in RCBs, while this value is higher by a
factor of 2 in case 4. The model F abundances are also spread over
nearly the whole range of observed values with the maximum model value
of $2.35\, \mathrm{dex}$ compared to the observed maximum of $2.7\,
\mathrm{dex}$. 

The net effect of the nucleosynthesis processes in the models is that
the $\mathrm{^{13}{C}}$ at the surface decreases while the
$\mathrm{^{12}{C}}$ increases (Section~3.2.2, Fig.\,\ref{part1} and
\ref{part2}), leading to an overall increase in the
$\mathrm{^{12}{C}/^{13}{C}}$ surface ratio. The
$\mathrm{^{12}{C}/^{13}{C}}$ ratios range from $\approx 230$ (case 4) to
$\approx 2500$ (case 3), much higher than the observed lower limit of
100.

The C, N, O and F model surface abundances rely exclusively on the
nucleosynthesis and mixing during the post-merger evolution
phase. Independent of the choice of the progenitor, the surface C, N
and O values of the four cases are within the same order
(Fig.\,\ref{x_mod}). The surface F abundances are close to each other
between cases 1, 2 and 4 (within a factor of 1.5,
Fig.\,\ref{o_f}). The highest F abundance is obtained in case 2 since
some of it was generated in the SOF by He burning, while the lowest
is in case 2 due to a much smaller depth of mixing, which led to a
smaller mass available to produce it. Case 4 has the highest amount of
O since its initial SOF was burnt at $2.42\times10^{8}$\,K, thereby
increasing the initial $^{16}\mathrm{O}$. This is reflected in the
high $^{16}\mathrm{O}$/$^{18}\mathrm{O}$ ratio $\sim$ 41 of this case
as well (Fig.\,\ref{o_f}). In summary, the model C, N, O and F abundances lie as
well in the observed range.

\subsubsubsection{Secondary elements}
\paragraph{Na - Ti:}

Both Na and Mg are enhanced compared to initial surface values in all four cases. Na is formed by proton captures on isotopes of Ne during the post-merger evolution. Thus, as the H-shell moves outwards in our simulations (Fig.\,\ref{part1}(i)a), the abundance of Na increases throughout the envelope. Na is also initially enriched in case 3 due to its production in the A-TP AGB progenitor used to construct the CO WD of this case.

For the rest of the species from Al to Ti, the abundances are
independent of the choice of the progenitor AGB model. In cases
1\,--\,3, these elements are changed in the low-$T$ SOF, depending on
its duration of burning, whereas in case 4 their abundances are
changed during the post-merger evolution phase (as mentioned in
Section~\ref{sec:sof}).




Al is first generated by proton-capture on Mg isotopes and later, being a neutron poison, is destroyed by (n,p) reactions once again to Mg via, $^{26}\mathrm{Al(n,p)}^{26}\mathrm{Mg}$. The Mg content also increases by proton-capture on $^{23}\mathrm{Na}$. Thus, although Mg is increased in all 4 cases, the level of reduction of Al depends on the interplay between the temperature and burning timescale in the SOF, which explains why the Al abundance can be either higher (cases 1 and 2) or lower than its initial value in the envelope (cases 3 and 4).

The abundance of Ca decreases in all cases, due to n-capture reactions
on its isotopes. $^{40}\mathrm{Ca}(\mathrm{n},\gamma)^{41}\mathrm{Ca}$
is a significant n-capture channel and the $^{41}\mathrm{Ca}$ thus
formed is used up by further (n,p) and (n,$\alpha$) reactions, thus
decreasing Ca. The reaction chain
$^{44}\mathrm{Ca}(\mathrm{n},\gamma)^{45}\mathrm{Ca}(\beta^{-})^{45}\mathrm{Sc}$
followed by
$^{45}\mathrm{Sc}(\mathrm{n},\gamma)^{46}\mathrm{Sc}(\beta^{-})^{46}\mathrm{Ti}$
further reduces Ca and in the process increases the amount of Ti.

From Fig.\,\ref{x_rcb}, the range of Na values from our models is
consistent with the data trend, while those of Al are in marginal
agreement with the enrichment levels seen in RCBs. Since there are
only 6 stars which have Mg measured and as they are well scattered, it
is difficult to draw out a trend in this element; we only barely get
the upper limits and do not get the decrements, as seen in two of
those stars. Upper limits are also closely achieved in Ti and Ca,
while our lower limits do not reach down to those found in some RCBs.

Ca is seen to be enriched in a few RCB stars with the lowest [Fe],
which none of our models show. Likewise, Si and S hardly change in all
four cases, as temperatures are not hot enough to produce them in the
initial SOF or in the star itself. Thus our models do not show the
overabundances of Si and S observed in RCB stars with the lowest [Fe]
(see discussion in Section~\ref{sec:disc}).

\paragraph{s-process elements:}
The abundances of s-process nuclei depend on the seed Fe nuclei
available in the star. Although the neutron source is primary, the
main neutron poison is also primary, and therefore the s~process is
secondary. Thus, one would expect [X] of s-process elements
to rise more or less in sync with [Fe].

Among our simulations the low-$q$ case with a high-$T$ SOF (case 4)
shows enhanced abundances of s-process elements from the neutron flux
available during the post-merger evolution phase (Section~\ref{sec:sof}), while the high-$q$ case with low-$T$ and long SOF
(case 2) produce the s-process elements in the SOF before the
post-merger evolution (Section~\ref{sec:sproc-pmerg}). Cases 1 and 3
produce only small amounts of s-process elements.

With few exceptions almost all RCB stars show some enhancement
in s-process elements  on their surface, compared to the solar-scaled
abundance level. The observed range of enhancements agrees with the
range of our four cases, except for a few
stars that seem to be depleted in s-process elements.


\section{Discussions and Conclusions}
\label{sec:disc}
Our post-merger models are based on realistic progenitor evolutions
with complete nucleosynthesis predictions and the dynamic merger
history presented in paper I. These models adopt an envelope mixing
model that is motivated by the rotation of the merged object. With
these assumptions the models reproduce almost all the observed
characteristics of RCB stars, in particular the anomalous isotopic
signatures of O, as well as C and the F enhancements.

Model predictions are also in good agreement with the observed
abundances of s-process elements. These do not originate in the
progenitor AGB star. Depending on the mass ratio of the
merging WDs the s process originates either in the SOF, which is a
characteristic of the dynamic phase (case 2) or in the post-merger
phase (case 4). In the former case s-process enhancements are only
seen if the SOF is assumed to persist for more than $10^6\mathrm{s}$
(long SOF).

The enhancements of C, N, O and F are due to nucleosynthesis in the
post-merger evolution of the star. The initial abundances constructed
from the progenitor AGB and RGB models are important with respect to the
amount of H available, and the profile of C and O into which dynamic and
post-merger dredge-up mixing may reach.  The amount of H is chosen
according to predictions of stellar evolution calculations of the giant
to white dwarf transition (paper I), and it provides the protons for the
neutron source isotope $^{13}\mathrm{C}$. Being of the order of 0.001 is
required to obtain the observed CNO elemental values of RCBs. 
Lower than this order of H, would cause the $^{14}\mathrm{N}$ abundance
to drop, and $\mathrm{C}>\mathrm{O}>\mathrm{N}$  while
higher than this causes $^{14}\mathrm{N}$ to rise above O at the surface, leading to
$\mathrm{C}>\mathrm{O}>\mathrm{N}$. The
majority of RCBs have $\mathrm{C}>\mathrm{N}>\mathrm{O}$ at the surface.

The CNO ratios will also depend on the abundance ratios in the outer
layer of the progenitor AGB stars core, and despite adopting a range
of progenitor AGB models, we find that the C, N, and O abundances are
similar for all cases.

Our models also lie within the observed range of luminosity and
effective temperature of RCB stars. The differences between the surface
abundances when the models enter and leave the RCB phase are
$<0.5\mathrm{dex}$ for C, N, O, and F, and $<0.2\mathrm{dex}$ for the
other nuclear species (this is less than the observational errors).
Thus, our models can also account for the four hot RCB stars between
$15000$ and $20000 \mathrm{K}$. The models show a range of
$^{16}\mathrm{O}/^{18}\mathrm{O}$ and F in agreement with
observations and their $\mathrm{^{12}{C}/^{13}{C}}$ ratios 
are higher than the observed lower limit for RCB stars.

The model parameters are constrained by the diverse range of
observational properties. In addition to the depth of mixing during
He-shell burning phase of the post-merger evolution ($k$), the mixing
limit $m_\mathrm{min}$ depends on the AGB progenitor. For an advanced
AGB progenitor model the depth of mixing must be limited to avoid an
excessive $\mathrm{^{16}O}$ dredge-up to the surface since the outer
layers of the evolving AGB CO-core get increasingly O-rich. Other
parameters of the mixing model (Section~\ref{sec:mixing}) calibrated to reproduce the
observed surface abundances are the width of the
partially mixed region below the proper mixing regime in the envelope
and the efficiency of mixing. The former needs to be large enough to
produce enough F but small enough to avoid destruction of
$\mathrm{^{18}O}$. The latter must be large enough to mix enough
$\mathrm{^{18}O}$ to the surface but small enough to avoid
contamination with $\mathrm{^{16}O}$. Surface abundances are also
determined by the SOF properties, which are mostly a result of the
mass ratio of the merging WDs. The duration of conditions for
nucleosynthesis in the SOF is uncertain and has been approximated by
assuming two cases, a long and short SOF. Finally, our models suggest
that the adopted additional envelope mixing should subside just before
the RCB star phase is reached. Without such a decrease of mixing
efficiency with time, continuous mixing throughout the evolution of
the star causes the star to become O-rich as against being C-rich.

Although in the end the choice of parameters (progenitor model, SOF
assumptions, mixing model) appears to have a lot of parameters, it
turns out that all of these seem to be identifiable with unique
nucleosynthetic and mixing signatures. Due to large number of
nucleosynthesis processes, each with their individual dependence on
temperature and on other, related nucleosynthesis processes (as for
example the neutron production and consumption for the F production),
most of these parameters are therefore well constrained in our models.

A legitimate question would be how degenerate our solutions are. In
view of the many cases that we have investigated, many of them not
described in detail in this paper, we are not aware of any alternative
way to arrive at such a level of agreement between observations and
models. However, one has to admit that significant uncertainties,
especially regarding the properties and evolution of the SOF as well
as non-spherically symmetric aspects of the post-merger mixing model
remain.

Observations of RCB stars show some diversity and our four cases
reproduce a spread in abundance features. Most of this spread in model
predictions is in fact due to differences in the cases, such as the
mass ratio of the WDs, or the progenitor model, which represent a real
source of diversity corresponding to the range of WD mergers that are
expected to end up as RCB or HdC stars. 

When comparing observations and model prediction it must be kept in
mind that dust depletion may play a role \citep{asplund98,asplund00}, since
after all the irregular variability of RCB stars is associated with
dust enshrouding. This possibility
cannot be ruled out,  in particular to explain the distinctly low [Fe]
values of the minority RCBs from the majority. It does not appear that
the dust-gas separation in RCB stars. An
ISM-like dust depletion can explain the high [S/Fe] and [Na/Fe]
observed in RCBs, but not their [Al/Fe] and [Ca/Fe] values, both of
which are much lower in the ISM. In particular, the Si in a
carbon-rich dust environment is expected to condense into SiC grains,
thus severely reducing the Si content. However all RCBs have high
[Si/Fe] ratios. It should be noted however, that only amorphous
graphite grains and no SiC grains have been detected in RCBs. This
maybe indicative of a different dust-gas separation mechanism from
that of the ISM.

An alternative interpretation of the large Si and S overabundances
seen in the RCB stars with the lowest [Fe] considers the
mass of the merging WDs. Si and S
are produced in O burning at $T \approx 10^{9}$\,K, which could be
achieved in the SOF of a merger of more massive WDs \citep[][$T=0.6 -
1.0 \times 10^{9}\,\mathrm{K}$ for combined mass $\geq
1.2M_\odot$]{loren-aguilar09}. The theoretical initial-final mass
relationship predicts, for a given initial mass, higher WD masses for
lower metallicity \citep{meng08}. Assuming the initial mass function does not depend
on metallicity in the range relevant for  RCB star progenitors, higher WD masses
are expected for lower metallicity mergers, and this may be the reason
for the larger enhancements in S and Si for the RCB stars with the
lowest Fe content. Such higher-mass RCB stars would naturally also be
more luminous, and this hypothesis therefore predicts that RCB star
luminosity is correlated with Si and S enhancements.

On a preliminary note, we do see enhancements of $^{22}\mathrm{Ne}$ on
the surface of our models that seem to agree with the levels of
$\mathrm{Ne}$ found in EHes. A detailed comparison of our models with
the observations of EHe is, however,  beyond the scope of
this work. 

Finally, a few exceptions to the usual RCB star properties
include those with higher than average H, including four RCB stars in which
Li is enhanced \citep{asplund00,kipper06}, as well as 
four other RCBs that have a relatively higher amount of $^{13}$C at
their surface \citep{rao08,asplund00}, with $^{12}$C/$^{13}$C$\lesssim
25$. Neither of these properties are found in our models. Recent
SPH simulation results of \citet{longland12} indicate that, to be
preserved, any Li that maybe produced in a DD merger will have to be
ejected into the mass that forms clouds around the RCB star. Since
mass-ejection during the dynamic merger phase is not part of the
hydrodynamic simulations of paper I, our models cannot predict lithium
abundances from such a scenario. Thus, the question of whether Li-rich
RCBs are an outcome of the FF scenario or of the merger between WDs
remains open and needs a separate, more in-depth analysis.

\acknowledgements This work has been supported, in part, by grant
NNX10AC72G from NASA's Astrophysics Theory Program. Herwig
acknowledges funding from NSERC through a Discovery Grant. This
research has been supported by the National Science Foundation under
grants PHY 11-25915 and AST 11-09174. This project was supported by
JINA (NSF grant PHY08-22648).  Pignatari thanks the support from an
Ambizione grant of the SNSF (Switzerland), and from Core project
Eurogenesis (MASCHE).  This project has made use of the MESA stellar
evolution code, and we acknowledge the contribution by the MESA
council, the MESA developers and the larger MESA user community,
without which the MESA code would not be the terrific research tool
that it is. This project has made use of the NuGrid suite of
nucleosynthesis codes, and we acknowledge the NuGrid collaboration
members for their support of the NuGrid platform.  This project has
made extensive use of NASA ADS and astro-ph.

%
%

\appendix
\section{Simulation data}
\label{sec:app_data}
In addition to the simulation results presented and described in this
paper we make all raw simulation output data from both the MESA
stellar evolution runs as well as the NuGrid mppnp post-processing
nucleosynthesis calculations available for further analysis or comparison with
observations or other simulations. The data is available via the NuGrid data server
\url{http://data.nugridstars.org} hosted at the Canadian Astronomy
Data Centre (CADC). Access information is available at
\url{http://www.nugridstars.org}. 

\end{document}